\documentstyle[12pt,psboxit]{cernman}
\def\z0{Z}
\def\gf{G_{\mu}}
\def\zm{M_{_Z}}

\def\tm{m_{t}}
\def\hm{M_{_H}}
\def\wm{M_{_W}}

\def\ge{\Gamma_{e}}
\def\gm{\Gamma_{\mu}}
\def\gt{\Gamma_{\tau}}
\def\gl{\Gamma_{l}}
\def\gu{\Gamma_{u}}
\def\gd{\Gamma_{d}}
\def\gc{\Gamma_{c}}
\def\gs{\Gamma_{s}}
\def\gb{\Gamma_{b}}
\def\gz{\Gamma_{_Z}}
\def\gh{\Gamma_{h}}

\def\afb{A_{_{FB}}}
\def\alr{A_{_{LR}}}

\def\alphahat{\hat \alpha}

\def\stes{\sin^2\theta}

\def\i3f{I_3^f}

\def\osp2{16\,\pi^2}
\def\ap2{\left(p^2\right)}

\def\Stt{\Sigma_{_{33}}}

\def\s0h{\sigma^0_h}

\begin{document}
\begin{titlepage}
\begin{flushright}
{preprint CPPM/95-1 \\[0.5cm]  ITEP/19-95 \\[2cm]}
\end{flushright}
\begin{center}
{\Huge \bf LEPTOP} \\[2cm]
\end{center}
{V.Novikov$^\ast$, L.Okun$^\ast$, A.Rozanov${^\star}$, and
 M.Vysotsky${^\ast}{^+}$\\[1cm]
$^\ast$ ITEP, Moscow 117259, Russia\\[0.5cm]
$^\star$  CPPM, IN2P3-CNRS, F-13288 Marseille, France\\[0.5cm]
$^+$ visitor at CPT CNRS and CPPM IN2P3-CNRS,
 F-13288 Marseille, France \\[2cm]}

\begin{center}
{\bf Abstract}
\end{center}

This report consists of three parts. In Chapter 1
we present a brief description of the  LEPTOP approach
for the calculation of
the radiative corrections in the Minimal Standard Model.
The approach is based on the one-loop approximation with respect to the
genuinely electroweak interaction for which simple explicit analytical
formulae are valid. Starting from $G_{\mu}, m_Z$ and $\bar{\alpha}\equiv
\alpha(m^2_Z)$ as the main input parameters,
we consider all observables of the $Z$ boson decays and the
mass of the $W$ boson on the same footing. The dressing with gluonic
corrections is performed by using the results of calculations
published by other authors.

In Chapter 2 theoretical uncertainties inherent to our approach
and hence to the LEPTOP computer program
are
analyzed.

In Chapter 3 we describe the Fortran code of LEPTOP.
 This code can be obtained
on request from rozanov@afsmail.cern.ch.

This document can be accessed on
 \Lit{http://cppm.in2p3.fr/leptop/intro_leptop.html}
by WWW users.

\end{titlepage}

\newpage
%\begin{center}
%\large{\bf Part I\\
%
%\vspace{3mm}
%
%LEPTOP BASIC EQUATIONS}
%\end{center}

\chapter{LEPTOP BASIC EQUATIONS}

\section{Introduction}

The program LEPTOP calculates radiative corrections to
the observables of
$Z$ boson decay and of the $W$ boson mass, $m_W$, in the Standard
 Model \cite{1}. It is based on a set of explicit formulae for these
observables, written in the papers \cite{2} -- \cite{9}.

Our aim is the study of the genuinely electroweak (e.-w.) interaction
by comparing theoretical results with experimental data. In
LEPTOP we separate  the e.-w. interaction
from the well known pure electromagnetic effects.
Such separation is well defined in the one loop e.-w. approximation.
Then we include all gluonic corrections (internal for hadron-free
observables and both internal and external for the $Z$ boson
hadronic decays) known in the literature.  The main result of our
study (refs. \cite{2}--\cite{9}) is that within $1\sigma$ --
$2\sigma$ accuracy experimental data are described by the e.-w. Born
approximation which is universal for any modification of the Minimal
Standard Model. The one-loop e.-w. corrections (that really specify the
e.-w. theory) are barely visible at the level of LEP1, SLD, UA2 and CDF
accuracy. We do not need to consider two-loop e.-w. corrections, except for
enhanced gaugeless top-quark contribution \cite{18a}-\cite{24b}.

In LEPTOP we confine  ourselves by considering effects larger than
$10^{-5}$.

\underline{Renormalization in LEPTOP}.  Loop calculations
depend on  parameter $\varepsilon = 4-D$ in
dimensional regularization  ($D$ is dimension of space-time),
 on the 't Hooft's
 scale
parameter $\mu$, on two bare
gauge coupling constants, vacuum expectation value of the Higgs field and on
the bare masses of quarks, leptons and
the Higgs boson.\footnote{In general they
also depend on Kobayashi--Maskawa angles but in the limit $m_q^2/m_Z^2\ll 1$
for all quarks $q\neq t$ this dependence vanishes for $Z$ boson decay and
$m_W$.} The dependence on $\varepsilon$ and $\mu$ disappears if one rewrites
 the
result of loop calculations in terms of renormalized coupling constants,
physical masses, etc., or in terms of corresponding number of some other
independent physical observables.

\underline{Input parameters}.
In LEPTOP we present the results of the loop calculations in terms of
three most accurately known electroweak parameters \cite{10},
\cite{11}:
\begin{equation}
m_Z = 91.1887(44) ~~ \mbox{\rm GeV}
\label{1}
\end{equation}
\begin{equation}
G_{\mu} = 1.16639(2)\cdot 10^{-5} ~~ \mbox{\rm GeV}^{-2}
\label{2}
\end{equation}
\begin{equation}
\bar{\alpha} \equiv \alpha(m_Z^2) = 1/128.87(12)
\label{3}
\end{equation}
and three free unknown parameters: physical top quark mass $m_t$ (on mass
shell), higgs  mass $m_H$ and strong coupling constant $\hat{\alpha}_s$
at scale $m_Z$: $\hat{\alpha}_s
\equiv \alpha_s (q^2 = m_Z^2)_{\overline{MS}}$; the latter enters
through gluonic corrections. There are tiny effects of nonzero
$b$-quark mass $m_b$, $\tau$-lepton mass $m_{\tau}$ and $c$-quark
mass $m_c$.  All other quarks and leptons can be considered as
massless.

Note that $G_{\mu}$, $\bar{\alpha}$ and $\hat{\alpha}_s$ are not directly
observable quantities:

1) $G_{\mu}$ is a  four-fermion coupling constant
of the muon decay corrected by purely
electromagnetic loops.

2) Electromagnetic coupling constant $\bar{\alpha}$ at scale $q^2 = m_Z^2$
includes running from the value $\alpha$ to $\bar{\alpha}$ due to
the three lepton
and five "light" quark loops calculated by formulae:
\begin{equation}
\bar{\alpha}=\frac{\alpha}{1-\delta \alpha}
\label{4}
\end{equation}
\begin{equation}
\delta \alpha = \frac{m_Z^2}{4\pi^2\alpha}
\int^{\infty}_{thr}\frac{ds}{m_Z^2
-s}\sigma_{e^+e^-}(s)
\label{5}
\end{equation}
where $\alpha$ is fine structure constant and $\sigma_{e^+e^-}$ is the cross
section of $e^+e^-$-annihilation, through one virtual photon, into leptons
and light hadrons:
\begin{equation}
\delta\alpha = \delta\alpha_l + \delta\alpha_h = 0.0314 + 0.0282(9)\;.
\label{301}
\end{equation}

The leptonic contribution
\begin{equation}
\delta\alpha_l = \frac{\alpha}{3\pi} \Sigma_l(\ln \frac{m^2_Z}{m^2_l} -
\frac{5}{3}) = 0.0314(0)
\label{302}
\end{equation}
is known to high accuracy. The hadronic contribution, $\delta\alpha_h$, is
calculated by using eq. (\ref{5}) with the experimental cross-section
$\sigma_{e^+e^- \to h}(s)$
in the dispersion integral below $s = (40 \mbox{\rm GeV})^2$
and parton result above $s
= (40 \mbox{\rm GeV})^2$ (see ref. \cite{11}).

3) The
strong coupling constant $\hat{\alpha}_s$ at scale $q^2 = m_Z^2$ is
defined as renormalized coupling in the
$\overline{MS}$ scheme of subtraction.

By using $\bar{\alpha}$ instead of  $\alpha$ we
automatically take into account the fact that the running weak coupling
constants $\alpha_W(q^2)$ and $\alpha_Z(q^2)$ do not run actually in the
region $|q^2|\leq m_Z^2$, while $\alpha(q^2)$ does run. So $\bar{\alpha}$ is
relevant to electroweak corrections (not $\alpha$ in spite of its
extremely high accuracy)
\cite{2}, \cite{7}.

The outline of Chapter 1 is as follows: In sect. 2 we introduce the general
notations  and phenomenological expressions for observables including the
final state interaction due to photon and gluon exchange up to the terms of
the order of $10^{-5}$. In sect. 3 we consider  the $\bar{\alpha}$-Born
approximation.  One e.-w. loop corrections are presented in
sect. 4 for hadron-free observables (sect. 4.1) and for $Z$ boson
decay into hadrons (sect. 4.2).  Chapter 2 is devoted to the accuracy
of the LEPTOP. In sect. 5 we make general remarks concerning various
sources of uncertainties. In sect. 6 we consider uncertainties in the
vector boson self-energies. Sect. 7 is devoted to the uncertainties
in the hadronic $Z$ decays. The procedure of estimating the total
theoretical accuracy of LEPTOP is described in sect. 8. Appendix A
presents the flowchart of LEPTOP.  Appendix B contains the full list
of our papers in which the LEPTOP approach was developed.

\section{Phenomenology and notations}

It is useful to define the electroweak angle $\theta \;\;(\sin\theta \equiv
s,\;\; \cos\theta
\equiv c)$ in terms of three basic parameters $G_{\mu}\;,\;\;
m_Z$ and $\bar{\alpha}$:
\begin{equation}
s^2c^2 = \frac{\pi\bar{\alpha}}{\sqrt{2}G_{\mu}m^2_Z}
\label{303}
\end{equation}
(for earlier references see \cite{12}). Solving eq. (\ref{303}) with
experimental values eqs. (\ref{1})--(\ref{3}) we get
\begin{eqnarray}
s^2 &=& 0.23117(33) \nonumber \\
c &=& 0.87683(19)
\label{304}
\end{eqnarray}

In LEPTOP the amplitude for
$Z$ boson decay into fermion-antifermion pair $f\bar{f}$ is written in one
e.-w. loop approximation in the following  form
\begin{eqnarray}
M(Z \to f\bar{f}) &=&
\frac{1}{2}\bar{f}Z_{\mu}\bar{\psi}_f(\gamma_{\mu}g_{Vf} + \gamma_{\mu}
\gamma_5 g_{Af})\psi_f \;\;,  \nonumber \\
\bar{f}^2 &=& 4\sqrt{2}G_{\mu}m^2_Z = 0.54866(8)
\label{6}
\end{eqnarray}
In this amplitude we neglect such structures as the  weak
magnetic moment induced by e.-w. interaction. These structures are of the
order of $(\alpha/\pi s^2)(m_f/m_Z)^2$, where $m_f$ is the fermion mass,
that is far beyond the accuracy of the present and future experiments; even
for $b$-quark it is $\sim 10^{-5}$.

By definition the constants $g_{V,A}$ do not include the contribution from
the interaction in the final state due to the exchange of gluons (for
quarks) and photons (for quarks and leptons). The final state interaction
includes also the processes of  photon and gluon
emission. Corresponding corrections have nothing to do
with electroweak corrections and can be written separately.

\underline{Leptonic widths}

For the decay into charged leptons $l\bar{l}$ we
explicitly take into account the final state QED
interaction factor
\begin{equation}
\Gamma_l =4\Gamma_0[(g^{l}_{V})^2
(1+\frac{3}{4\pi}\bar{\alpha})+(g^{l}_{A})^2
(1+\frac{3}{4\pi}\bar{\alpha}-6\frac{m^2_l}{m^2_Z})]
%)4(g^2_{Al}+-6g^2_{Al}
%)  \;\;,
\label{7}
\end{equation}
where
\begin{equation}
\Gamma_0 = \frac{1}{24\sqrt{2}\pi} G_{\mu}m^3_Z = 82.945(12)
 \mbox{\rm MeV}\;\;.
\label{8}
\end{equation}
Second order QED corrections  $(\frac{\bar{\alpha}}{\pi})^2
\sim 10^{-6}$ are neglected. The term proportional to leptonic
 mass squared in
 eq.
(\ref{7}) is also negligible
 for $l = e, \mu$ (so we put $m_e = m_{\mu} = 0$) and is barely visible
only for $l = \tau$ $(m^2_{\tau}/m^2_Z = 3.8 \cdot 10^{-4})$.

For the neutrino decay
\begin{eqnarray}
\Gamma_{\nu} &=& 8g^2_{\nu} \Gamma_0\;\; ,  \nonumber  \\
g_{\nu} &=& g_{V\nu} = g_{A\nu}\;\; .
\label{9}
\end{eqnarray}

\underline{Hadronic widths}

For the decay into light quarks $q = u,d,s$ we neglect small effects due
to nonzero quark mass (i.e. we put $m_u = m_d = m_s = 0$) and take
into account the final state
 gluon exchange  up to the
third order \cite{13}, \cite{14}, \cite{20}, \cite{21},
the final state one photon exchange and the  interference of the photon
and gluon exchange \cite{100}.
These corrections
are  slightly
different for vector and axial channels.

For the decay into quarks we have
\begin{equation}
\Gamma_q = \Gamma(Z \to q\bar{q}) = 12[g^2_{Aq} R_{Aq} + g^2_{Vq}
R_{Vq}]\Gamma_0
\label{10}
\end{equation}
where factors $R_{A,V}$ are due to
the final state interaction. (In our previous papers we used letter $G$
instead of $R$ for these factors).  According to \cite{14}, \cite{20},
\cite{21}
\begin{eqnarray}
R_{Vq} &=& 1 + \frac{\hat{\alpha}_s}{\pi} +
\frac{3}{4} Q^2_q \frac{\bar{\alpha}}{\pi} - \frac{1}{4} Q^2_q
\frac{\bar{\alpha}}{\pi} \frac{\hat{\alpha}_s}{\pi} +  \nonumber  \\
&+&
[1.409 + (0.065 + 0.015 \ln t)\frac{1}{t}](\frac{\hat{\alpha}_s}{\pi})^2
-12.77(\frac{\hat{\alpha}_s}{\pi})^3
+12 \frac{\hat{m}^2_q}{m^2_Z} \frac{\hat{\alpha}_s}{\pi} \delta_{vm}
\label{11}
\end{eqnarray}
%\begin{eqnarray}
$$R_{Aq} = R_{Vq} - (2T_{3q})[I_2(t)(\frac{\hat{\alpha}_s}{\pi})^2 +
I_3(t)(\frac{\hat{\alpha}_s}{\pi})^3]
-$$
\begin{eqnarray}
-12 \frac{\hat{m}^{2}_{q}}{m^{2}_{Z}} \frac{\hat{\alpha}_s}{\pi}
 \delta_{vm} -
6\frac{\hat{m}^2_q}{m^2_Z} \delta^1_{am}
- 10 \frac{\hat{m}^2_q}{m^2_t}(\frac{\hat{\alpha}_s}{\pi})^2 \delta^2_{am}
\label{12}
\end{eqnarray}
where $\hat{m}_q$ is running quark mass (see below),
\begin{eqnarray}
\delta_{vm} = 1+8.7
(\frac{\hat{\alpha}_s}{\pi})+45.15(\frac{\hat{\alpha}_s}{\pi})^2,
\label{11101}
\end{eqnarray}
\begin{eqnarray}
\delta^1_{am} = 1 + 3.67(\frac{\hat{\alpha}_s}{\pi}) +(11.29 - \ln t)
(\frac{\hat{\alpha}_s}{\pi})^2,
\label{11102}
\end{eqnarray}
\begin{eqnarray}
\delta^2_{am} = \frac{8}{81}+\frac{\ln t}{54},
\end{eqnarray}
\begin{equation}
I_2(t) = -3.083 - \ln t + \frac{0.086}{t} + \frac{0.013}{t^2}\;\;,
\label{13}
\end{equation}
\begin{eqnarray}
I_3(t) &=& -15.988 - 3.722 \ln t + 1.917 \ln^2 t\;\;, \\ \nonumber
t &=& m^2_t/m^2_Z\;\;.
\label{14}
\end{eqnarray}
The terms of the order of $(\frac{\hat{\alpha}_s}{\pi})^3$ due to diagrams
with three gluons in the intermediate state have been calculated for
$R_{Vq}$ in ref.\cite{21}.
They have small numerical coefficients so that the net effect is of the
order of $10^{-5}$ and we have omitted these terms in expansion
(\ref{11}).\\
For $Z \to b\bar{b}$ decay the nonzero $b$-quark mass is not negligible and
produces contribution of the order of 1 MeV to
$\Gamma_b$ (i.e. of the order of 0.5--0.3\%). Gluonic corrections
effectively change the pole mass $m_b \simeq 4.7 \mbox{\rm GeV}$
 to the running mass
at scale $m_Z\;:\;\; m_b \to \hat{m}_b(m_Z)$.
We calculate $\hat{m}_b$ in terms of pole mass $m_b\;,\;\;
\hat{\alpha}_s(m_Z)$ and $\hat{\alpha}_s(m_b)$ using standard two loops
equation in $\overline{MS}$ scheme (see e.g. ref. \cite{101}).\\
For $Z \to c\bar{c}$ decay the running mass $\hat{m}_c(m_Z)$
 is expected to be
of the order of 0.5 GeV and corresponding contribution  to $\Gamma_c$ is of
the order of $0.05$ MeV. Actually we take into account this tiny effect in
LEPTOP because it is included in the other computer codes (ZFITTER).\\
In connection with $\Gamma_c$, let us note that the term $I_2(t)$ given by
eq. (\ref{13}) contains interference terms which appear at the order
$(\frac{\hat{\alpha} s}{\pi})^2$,  hence its negative sign. These terms have
three types of final states: one quark pair, one quark pair plus gluon, two
quark pairs. The last final state constitutes about 5 \% of the $I_2$ and
at the present level of experimental accuracy is negligibly small. However,
in principle, it calls for special care, especially when the two pairs
consist of quarks of different flavor, e.g.  $b\bar{b}c\bar{c}$. Such mixed
final states should be a subject of special negotiations between  theorists
and experimentalists.

\underline{Asymmetries}

a) Forward-backward asymmetry into $f\bar{f}$ channel is given by the
equation
\begin{equation}
A^{f\bar{f}}_{FB} = \frac{3}{4} A_e A_f
\label{18}
\end{equation}
where for light fermions
\begin{equation}
A_f = \frac{2g_{Af} g_{Vf}}{(g_{Af})^2 + (g_{Vf}
)^2}\;\;.
\label{19}
\end{equation}
For $b$-quark we take into account the effect of nonzero mass:
\begin{equation}
A_b = \frac{2g_{Ab} g_{Vb}}{[v^2g^2_{Ab} + \frac{1}{2}(3-v^2)g^2_{Vb}]} v \;
,
\label{20}
\end{equation}
where $v$ is the $b$-quark velocity:
\begin{equation}
v = \sqrt{1 - \frac{4 m^2_b(m^2_Z)}{m^2_Z}}\;\;.
\label{305}
\end{equation}
As was already mentioned,
it is impossible to separate a given quark channel from another one
unambiguously,
starting from the order $(\frac{\hat{\alpha}_s}{\pi})^2$,
just due to the possibility of creating an additional pair of
"foreign" quarks. So we prefer not to consider terms of the order of
$(\frac{\alpha_s}{\pi})^2$ in asymmetries at all. In this approximation the
ratio $g_V/g_A$ is not renormalized by the final state interaction.\\
b) Longitudinal polarization of $\tau$-lepton
\begin{equation}
P_{\tau} = -A_{\tau}
\label{21}
\end{equation}
c) The relative difference of total cross-section at $Z$ peak for left- and
right-handed electrons colliding with unpolarized positron beam
\begin{equation}
A_{LR} = \frac{\sigma_L - \sigma_R}{\sigma_L + \sigma_R} = A_e
\label{22}
\end{equation}

Other observables are defined in terms of $\Gamma_q$ and $\Gamma_l$ by
equations:

a) Hadronic width (up to very small
corrections) is given by the sum of five quark
channels:
\begin{equation}
\Gamma_h = \Gamma_u + \Gamma_d + \Gamma_c +
\Gamma_s + \Gamma_b \label{23}
\end{equation}

b) Total width:
\begin{equation}
\Gamma_Z = \Gamma_h + \Gamma_e + \Gamma_{\mu} + \Gamma_{\tau} +
3\Gamma_{\nu}
\label{24}
\end{equation}

c) Peak cross section of $e^+e^-$-annihilation into hadrons:
\begin{equation}
\sigma_h = \frac{12\pi}{M^2_Z} \frac{\Gamma_e\Gamma_h}{\Gamma^2_Z}
\label{25}
\end{equation}

d) Ratios
\begin{equation}
R_c = \frac{\Gamma_c}{\Gamma_h}~,~~~R_b = \frac{\Gamma_b}{\Gamma_h}~,
{}~~~ R_l = \frac{\Gamma_h}{\Gamma_l}~~~.
\label{26}
\end{equation}

\section{The $\bar{\alpha}$--Born approximation}

If we write the expressions for observables  in the Born
approximation in terms of $s$ and $c$  we
automatically include purely electromagnetic interaction. Thus we get:
\begin{equation}
(m_W/m_Z)^B = c
\label{28}
\end{equation}
\begin{equation}
(g_{Af})^B = T_{3f}
\label{29}
\end{equation}
\begin{equation}
(g_{Vf}/g_{Af})^B = 1-4|Q_f|s^2
\label{30}
\end{equation}

Substituting $(g_{V,A})^B$ into equations for
the widths and asymmetries from the
previous section we calculate the whole set of observables in $Z$ decay in
the Born approximation for electroweak interaction, but take into account
trivial pure electromagnetic and gluonic interaction in the final state and
the effect of the running of e.-m. coupling constant. Rather luckily for
LEPTOP
these equations reproduce
the precision experimental data for $Z$ decay and for
$m_W$ (Jan. 95) with the accuracy $1\sigma - 1.5 \sigma$ \cite{6}, \cite{7},
\cite{15}. (Note that the final state interaction  factors (eqs. (\ref{11})--
(\ref{12})) include a lot of  very small terms. One can neglect mass terms,
as well as the difference between  the vector
and axial channels eq.  (\ref{12}). Such simplified Born
approximation will still reproduce the data with practically the same
accuracy).

It is important that the Born approximation for "low"--energy observables
(parameters  of $Z$ boson decay and $m_W$) is the same for very different
models, hence the present experimental confirmation of the
$\bar{\alpha}$-Born approximation
is not sufficient
to prove the validity of the MSM or to choose between different
generalizations.

\section{Electroweak loop corrections}
\subsection{Hadron-free observables}

For hadron-free observables we write the result of one-loop e.-w.
calculations in the form suggested in ref. \cite{2}:
\begin{eqnarray}
m_W/m_Z = c+\frac{3c}{32\pi s^2(c^2-s^2)}\bar{\alpha}V_m(t,h) \nonumber \\
g_{Al} = -1/2-\frac{3}{64\pi s^2c^2}\bar{\alpha}V_A(t,h) \nonumber \\
\\
R =
g_{Vl}/g_{Al} =
1- 4s^2 + \frac{3}{4\pi(c^2 - s^2)}\bar{\alpha}V_R(t,h)
\nonumber
\\
g_{\nu} = 1/2+ \frac{3}{64\pi s^2c^2}\bar{\alpha}V_{\nu}(t,h)
\nonumber
\label{31}
\end{eqnarray}
where $t=m_t^2/m_Z^2$, $h= m_H^2/m_Z^2$
and functions $V_i(t,h)$ are normalized by the condition
\begin{equation}
V_i(t,h)\simeq t
\label{32}
\end{equation}
at $t\gg 1$.

Each function $V_i$ is a sum of five terms
\begin{equation}
V_i(t,h) = t + T_i(t)+H_i(h)+C_i+\delta V_i(t,h)
\label{33}
\end{equation}

The functions $t + T_i(t)$ are due to  $(t,b)$ doublet contribution
to self-energies  of the vector bosons, $H_i(h)$ is due to $W^{\pm},Z$ and
$H$ loops, the constants $C_i$ include light fermion contribution both to
self energies, vertex and box diagrams.

To give explicit expressions for $T_i(t)$ and $H_i(h)$ it is convenient to
introduce three auxiliary functions $F_t(t)$, $F_h(h)$ and $F'_h(h)$ (see
subsection 4.3  for their expressions).

The equations for $T_i(t)$ and $H_i(h)$ have the form \cite{2}:
$$
\underline{i = m}
$$
$$
T_m(t) = (\frac{2}{3} - \frac{8}{9}s^2)\ln t - \frac{4}{3} + \frac{32}{9}
s^2 +
$$
$$
+ \frac{2}{3}(c^2 - s^2)(\frac{t^3}{c^6} - \frac{3t}{c^2} + 2) \ln \mid 1 -
\frac{c^2}{t} \mid +
$$
\begin{equation}
+ \frac{2}{3} \frac{c^2-s^2}{c^4} t^2 + \frac{1}{3} \frac{c^2 - s^2}{c^2} t
+ [\frac{2}{3} - \frac{16}{9} s^2 -
\frac{2}{3} t - \frac{32}{9} s^2t] F_t(t)\;;
\label{34}
\end{equation}
$$
H_m(h) = -\frac{h}{h-1} \ln h + \frac{c^2 h}{h-c^2}\ln \frac{h}{c^2} -
\frac{s^2}{18c^2} h - \frac{8}{3}s^2 +
$$
$$
+ (\frac{h^2}{9} -\frac{4h}{9} + \frac{4}{3})F_h(h) -
$$
$$
-(c^2 - s^2)(\frac{h^2}{9c^4} - \frac{4}{9} \frac{h}{c^2} + \frac{4}{3})
F_h(\frac{h}{c^2}) +
$$
$$
+ (1.1205 - 2.59\delta s^2)\;;
$$
where $\delta s^2 = 0.23117 - s^2$ (note the sign!).
$$
\underline{i = A}
$$
$$
T_A(t) = \frac{2}{3} - \frac{8}{9} s^2 + \frac{16}{27} s^4 - \frac{1 -
2tF_t(t)}{4t-1} +
$$

\begin{equation}
+(\frac{32}{9} s^4 - \frac{8}{3} s^2 - \frac{1}{2})[ \frac{4}{3}tF_t(t) -
\frac{2}{3}(1+2t)\frac{1-2tF_t(t)}{4t-1}]
\label{35}
\end{equation}
$$
H_A(h) = \frac{c^2}{1-c^2/h}\ln \frac{h}{c^2} - \frac{8h}{9(h-1)}\ln h +
$$
$$
+(\frac{4}{3} - \frac{2}{3}h+ \frac{2}{9}h^2)F_h(h) - (\frac{4}{3} -
\frac{4}{9}h + \frac{1}{9} h^2) F'_h(h)-
$$
$$
-\frac{1}{18}h + [0.7751 + 1.07\delta s^2]
\;.$$

$$
\underline{i = R}
$$
$$
T_R(t) = \frac{2}{9}\ln t + \frac{4}{9} -
\frac{2}{9}(1+ 11t)F_t(t)\;;
$$
\begin{equation}
H_R(h) = -\frac{4}{3} - \frac{h}{18} + \frac{c^2}{1 - c^2/h}\ln
\frac{h}{c^2} +
\label{36}
\end{equation}
$$
+ (\frac{4}{3} - \frac{4}{9}h + \frac{1}{9}h^2)F_h(h) + \frac{h}{1-h}\ln h +
$$
$$
+ (1.3590 + 0.51\delta s^2)
$$
$$
\underline{i=\nu}
$$
$$
T_{\nu}(t) = T_A(t)
$$
\begin{equation}
H_{\nu}(h) = H_A(h)
\label{37}
\end{equation}

The constants  $C_i$ are rather complicated  functions of $\sin^2 \theta$
and we present their numerical values at $s^2 = 0.23117- \delta s^2$
\begin{equation}
C_m = -1.3500 + 4.13 \delta s^2
\label{501}
\end{equation}
\begin{equation}
C_A = -2.2619 - 2.63 \delta s^2
\label{502}
\end{equation}
\begin{equation}
C_R = -3.5041 - 5.72 \delta s^2
\label{503}
\end{equation}
\begin{equation}
C_{\nu} = -1.1638 - 4.88\delta s^2
\label{504}
\end{equation}

Functions $\delta V_i(t,h)$ in eq. (\ref{33}) are small corrections to
$V_i$.  They can be separated into five classes:

\begin{enumerate}
\item{Corrections due to $W$-boson $\delta_W\alpha$ and $t$-quark
$\delta_t\alpha$
polarization of e.-m. vacuum are traditionally not included into
the running of $\alpha(q^2)$. We
also prefer to consider them together with
electroweak corrections. This is especially reasonable because $W$
contribution $\delta_W\alpha$ is gauge dependent. Here and for all other
e.-w.  corrections we use 't Hooft--Feynman gauge. The corrections
$\delta_W\alpha$ and $\delta_t\alpha$ were neglected in ref. \cite{2} and
were introduced in ref. \cite{7}.
\begin{equation}
\delta_1 V_m(t,h) =
-\frac{16}{3} \pi s^4 \frac{1}{\alpha}(\delta_W\alpha + \delta_t \alpha)
= -0.055 \;\; ,
\label{405}
\end{equation}
\begin{equation}
\delta_1 V_R(t,h) = -\frac{16}{3} \pi s^2 c^2
\frac{1}{\alpha}(\delta_W\alpha + \delta_t\alpha) = -0.181 \;\; ,
\label{38}
\end{equation}
\begin{equation}
\delta_1 V_A(t,h) = \delta_1 V_{\nu}(t,h) = 0\;,
\label{306}
\end{equation}
where
\begin{equation}
\frac{\delta_W\alpha}{\alpha} = \frac{1}{2\pi} [(3 + 4c^2)(1
-\sqrt{4c^2-1} \arcsin \frac{1}{2c}) - \frac{1}{3}] = 0.0686 \; ,
\label{307}
\end{equation}
\begin{equation}
\frac{\delta_t\alpha}{\alpha} = -\frac{4}{9\pi}[(1+2t)F_t(t) - \frac{1}{3}]
\simeq - \frac{4}{45\pi}\frac{1}{t} + ... \simeq
- 0.00768 \;\; .
\label{308}
\end{equation}
(Here and in the equations (\ref{309}) -- (\ref{319}) we use $m_t = 175$
GeV for numerical estimates.)}

\item{Corrections of the order of
$\bar{\alpha}\hat{\alpha}_s$ due to the gluon exchange
in the quark e.-w. loops \cite{16} (see also \cite{3}). For the two
generations of light quarks $(q = u,d,s,c)$ it gives:
\begin{equation}
\delta^q_2 V_m(t,h)
= 2 \cdot [\frac{4}{3}(\frac{\hat{\alpha}_s(m_Z)}{\pi})(c^2-s^2)\ln c^2] =
(\frac{\hat{\alpha}_s(m_Z)}{\pi})(-0.377)
\label{309}
\end{equation}
\begin{equation}
\delta^q_2 V_A(t,h) = \delta^q_2 V_\nu(t,h) = 2 \cdot
 [\frac{4}{3}(\frac{\hat{\alpha}_s(m_Z)}{\pi})
(c^2-s^2 + \frac{20}{9} s^4)]=
(\frac{\hat{\alpha}_s(m_Z)}{\pi})(1.750)
\label{310}
\end{equation}
\begin{equation}
\delta^q_2 V_R(t,h) = 0
\label{311}
\end{equation}

The result of calculation for third generation is rather complicated
function of top-mass:
$$
 \delta^t_2 V_m(t,h) = \frac{4}{3} (\frac{\hat{\alpha}_s (m_t)}{\pi})
\{tA_1 (\frac{1}{4t})+(1-\frac{16}{3}s^2)tV_1(\frac{1}{4t})
+(\frac{1}{2}-\frac{2}{3}s^2) \ln t
$$
\begin{equation}
-4(1-\frac{s^2}{c^2}) \times tF_1(\frac{c^2}{t})-4\frac{s^2}{c^2}tF_1 (0)\}
\label{11103}
\end{equation}
$$
\delta ^t_2 V_A (t,h) = \delta ^t_2 V_\nu (t,h) =
 \frac{4}{3}(\frac{\hat{\alpha}_s (m_t)}{\pi})
\{tA_1(\frac{1}{4t})-\frac{1}{4}A^{'}_{1} (\frac{1}{4t})+
$$
\begin{equation}
+(1-\frac{8}{3}s^2 )^2[tV_1 (\frac{1}{4t})-
\frac{1}{4}V^\prime _1 (\frac{1}{4t})]+(\frac{1}{2}-\frac{2}{3}s^2 +
 \frac{4}{9}s^4 )-4tF_1 (0)\}
\label{11104}
\end{equation}

\begin{equation}
\delta^t_2 V_R(t,h) = \frac{4}{3} (\frac{\hat{\alpha}_s (m_t)}{\pi})\{tA_1
 (\frac{1}{4t})
-\frac{5}{3}tV_1 (\frac{1}{4t})-4tF_1(0)+\frac{1}{6} \ln t\},
\label{11105}
\end{equation}
where
\begin{equation}
\hat{\alpha}_s (m_t) =
\frac{\hat{\alpha}_s (\zm)}
{1 +\frac{23}{12 \pi}{\hat{\alpha}_s (\zm)} \ln t}
\label{11105a}
\end{equation}

Note that $\delta _2 V_i $, unlike $V_i$ themselves, do not depend on $m_H$.
 The
 functions
$V_1 (r)$, $A_1 (r)$ and $F_1 (x)$ have a very complicated form and were
 calculated in ref. \cite{16}.
For our purpose we can use their rather simple Taylor expansion for
 small values
 of their arguments
(we have added cubic terms to the expansion presented in ref. \cite{16}):

\begin{equation}
V_1 (r) = r[4\zeta(3)-\frac{5}{6}]+r^2 \frac{328}{81}+r^3
 \frac{1796}{25 \times
 27}+...
\label{11106}
\end{equation}

\begin{equation}
A_1 (r) = [-6\zeta (3)-3\zeta (2) + \frac{21}{4}] + r[4\zeta (3) -
 \frac{49}{18}]+
r^2 \frac{689}{405} + r^3 \frac{3382}{7 \times 25 \times 27} + ...
\label{11107}
\end{equation}

$$
F_1 (x)=[-\frac{3}{2} \zeta (3) - \frac{1}{2}\zeta (2) +
 \frac{23}{16}]+x[\zeta
 (3)-
\frac{1}{9} \zeta (2) - \frac{25}{72}]+
$$
\begin{equation}
+x^2[\frac{1}{8}\zeta (2) + \frac{25}{3 \times 64}]+x^3
[\frac{1}{30} \zeta (2) +\frac{5}{72}]+ ...
\label{11108}
\end{equation}
where $\zeta (2) = \pi ^2 /6, \zeta (3) = 1.2020569 ....$

By summing contributions of (\ref{11103})-(\ref{11105}) and using expansions
 (\ref{11106})-(\ref{11108})
one obtains up to terms $0(1/t^3)$:

\begin{equation}
\delta^t_2 V_m(t,h) = (\frac{\hat{\alpha}_s(m_t)}{\pi})[-2.86t + 0.46\ln t -
 1.540 - \frac{0.68}{t} - \frac{0.21}{t^2}] =
\frac{\hat{\alpha}_s(m_t)}{\pi}(-11.67)
\label{312}
\end{equation}
\begin{equation}
\delta^t_2 V_A(t,h) = \delta^t_2 V_\nu (t,h) =
(\frac{\hat{\alpha}_s(m_t)}{\pi})[-2.86t + 0.493 - \frac{0.19}{t} -
\frac{0.05}{t^2}]=\frac{\hat{\alpha}_s(m_t)}{\pi}(-10.10)
\label{313}
\end{equation}
\begin{equation}
\delta^t_2
V_R(t,h) = (\frac{\hat{\alpha}_s(m_t)}{\pi})[-2.86t + 0.22 \ln t - 1.513 -
\frac{0.42}{t} - \frac{0.08 }{t^2}]=\frac{\hat{\alpha}_s(m_t)}{\pi}(-11.88)
\label{314}
\end{equation}
As these formulas are valid for $m_t > m_Z$, in order
to go to the region $m_t < m_Z$ we either put
$\delta^t_2 V_i = 0$ or use massless limit in which $\delta^t_2 V_i =
\frac{1}{2}\delta^q_2 V_i$. In any case this region gives tiny
contribution into global fit.}
\item{Corrections of the order of
$\bar{\alpha}\hat{\alpha} ^2_s$ were calculated for leading term
$\bar{\alpha} \hat{\alpha}^2_s t$ only \cite{17}
\begin{equation}
\delta_3 V_i(t,h) \simeq -(2.38 - 0.18 N_f) \hat{\alpha}^2_s(m_t)t
\simeq -1.48 \hat{\alpha}^2_s (m_t)t = -0.07
\label{315}
\end{equation}
for $N_f = 5$ light flavors. (For numerical estimate we use
$\hat{\alpha}_s(m_Z)=0.125$.)}
\item{The leading correction of the order of
$\bar{\alpha}^2 t^2$ which originates from
the second order Yukawa interaction
was calculated in ref.  \cite{18a}-\cite{24b}
\begin{equation}
\delta_4 V_i(t,h) =
-\frac{\bar{\alpha}}{16\pi s^2c^2} A(\frac{h}{t}) \cdot t^2
\label{316}
\end{equation}

Function $A(\frac{h}{t})$ is given in the table \ref{tab1}
\cite{24a}-\cite{24b} for $m_H/m_t < 4$. We use expansion
 from \cite{18a}-\cite{18b} for $m_H / m_t >4$:
$$
A(\frac{h}{t}) = \frac{49}{4} + \pi^2 + \frac{27}{2} \ln r +
 \frac{3}{2} \ln^2 r + \frac{1}{3}r(2-12\pi^2 +12\ln r-27\ln^2 r) +
$$
\begin{equation}
+\frac{1}{48}r^2(1613 - 240\pi^2 -1500\ln r-720\ln^2 r),
\label{316a}
\end{equation}
 where $r=t/h$.
  For $m_t = 175$ GeV and $m_H = 300$ GeV one has
$A = 8.9$ and $\delta_4V_i(t,h) = -0.11$.}

\item{In the second order in e.-w. interactions there appears quadratic
dependence on the higgs mass \cite{19}.
\begin{equation}
\delta_5 V_m
=\frac{\bar{\alpha}}{24\pi}(\frac{m_H^2}{m_Z^2})\times \frac{0.747}{c^2}
 = 0.0011
\label{317}
\end{equation}
\begin{equation}
\delta_5 V_A = \delta_5 V_\nu
=\frac{\bar{\alpha}}{24\pi}(\frac{m_H^2}{m_Z^2})\times \frac{1.199}{s^2}
= 0.0057
\label{318}
\end{equation}
\begin{equation}
\delta_5 V_R =-\frac{\bar{\alpha}}{24\pi}(\frac{m_H^2}{m_Z^2})\frac{c^2 -s^2}
{s^2 c^2}\times 0.973 = -0.0032 \;\; .
\label{319}
\end{equation}

(For numerical estimates we use $m_H = 300$ GeV.)}
\end{enumerate}

\subsection{$Z$ boson decay into hadrons}

For the partial width of the $Z$ decay into a pair of quarks
$q\bar{q}$ ($q = u, d, s, c, b$) we use the equation (\ref{10})
$$
\Gamma_q = 12[g^2_{Aq}R_{Aq} +g^2_{Vq}R_{Vq}]\Gamma_0 \;\; ,
$$
where e.-w. radiative corrections are included into $g_{Vq}$ and $g_{Aq}$
\begin{equation}
g_{Aq} = T_{3q}[1+\frac{3\bar{\alpha}}{32\pi s^2 c^2}V_{Aq}(t,h)]
\label{320}
\end{equation}
\begin{equation}
g_{Vq}/g_{Aq} = 1-4|Q_q|s^2 +\frac{3|Q_q|}{4\pi(c^2
-s^2)}\bar{\alpha}V_{Rq}(t,h) \;\; .
\label{321}
\end{equation}
The functions $V_{Aq}(t,h)$ and $V_{Rq}(t,h)$ in the one-loop e.-w-
approximation are related to the functions $V_A(t,h)$ and $V_R(t,h)$
from leptonic decays \cite{5}:
%$$
\begin{equation}
V_{Au}(t,h) = V_{Ac}(t,h) = V_{A}(t,h) + \frac{128\pi s^3 c^3}{3 \bar{\alpha}}
 (F_{Al}+F_{Au})
\label{322}
\end{equation}

%$$
\begin{equation}
V_{Ad}(t,h) = V_{As}(t,h) = V_{A}(t,h) + \frac{128\pi s^3 c^3}{3 \bar{\alpha}}
 (F_{Al}-F_{Ad})
%$$
\label{323}
\end{equation}
%$$
\begin{eqnarray}
V_{Ru}(t,h) = V_{Rc}(t,h) = V_{R}(t,h)+ \frac{16\pi sc(c^2 - s^2)}{3
 \bar{\alpha}} \times \\ \nonumber
 \times [F_{Vl}-(1-4s^2)F_{Al} + \frac{3}{2}(-(1-\frac{8}{3}s^2 )F_{Au}
+F_{Vu})]
\label{324}
\end{eqnarray}
%$$
\begin{eqnarray}
V_{Rd}(t,h) = V_{Rs}(t,h) = V_{R}(t,h)+ \frac{16\pi sc(c^2 - s^2)}{3
 \bar{\alpha}} \times \\ \nonumber
\times [F_{Vl}-(1-4s^2)F_{Al} + 3((1-\frac{4}{3}s^2 )F_{Ad}-F_{Vd})]
\label{325}
\end{eqnarray}
where:
\begin{equation}
F_{Al} = \frac{\bar{\alpha}}{4\pi}(3.0088 + 16.4 \delta s^2)\;,
\label{741}
\end{equation}
\begin{equation}
F_{Vl} = \frac{\bar{\alpha}}{4\pi}(3.1868 + 14.9 \delta s^2)\;,
\label{742}
\end{equation}
\begin{equation}
F_{Au} = - \frac{\bar{\alpha}}{4\pi}(2.6792 + 14.7 \delta s^2)\;,
\label{743}
\end{equation}
\begin{equation}
F_{Vu} = - \frac{\bar{\alpha}}{4\pi}(2.7319 + 14.2 \delta s^2)\;,
\label{744}
\end{equation}
\begin{equation}
F_{Ad} =  \frac{\bar{\alpha}}{4\pi}(2.2212 + 13.5 \delta s^2)\;,
\label{745}
\end{equation}
\begin{equation}
F_{Vd} =  \frac{\bar{\alpha}}{4\pi}(2.2278 + 13.5 \delta s^2)\;.
\label{746}
\end{equation}

The five digit accuracy of the above numbers is purely arithmetical.
The physical uncertainties (caused by the neglect of the higher loop
corrections) are at the level of the third digit.

The difference $V_{iq} - V_i$ (where $i=A,R$) is due to different
e.-w.  corrections to the vertices $Zq\bar{q}$ and $Zl\bar{l}$.

The oblique corrections of the order of $\hat{\alpha}_s$,
$\hat{\alpha}_s^2 t$ and $\bar{\alpha} t^2$
 to the $V_{Aq}(V_{Rq})$ are the same
as in the case of $V_A$ and $V_R$. But for $Z$-boson decay into pair
$q\bar{q}$ there are also additional $\hat{\alpha}_s$ corrections to the
vertices that have not been calculated yet.
That brings additional uncertainty
into the theoretical accuracy.

For $Z\to b\bar{b}$ decays we have to take into account
 corrections to the $Z\to
 b\bar{b}$ vertex which
depend on $t$ \cite{22}, \cite{23}

\begin{equation}
V_{Ab}(t,h) =
 V_{Ad}(t,h)-\frac{8s^2 c^2}{3(3-2s^2)}(\phi (t) + \delta \phi (t))
 ,
\label{11109}
\end{equation}
\begin{equation}
V_{Rb}(t,h) = V_{Rd}(t,h)-\frac{4s^2 (c^2 -s^2 )}{3(3-2s^2)}(\phi (t) + \delta
 \phi (t))
\label{11110}
\end{equation}
where for $\phi(t)$ we use the following expansion \cite{22}
$$
\phi(t) = \frac{3-2s^2}{2s^2 c^2}\left\{t+c^2[2.88 ln\frac{t}{c^2} -6.716 +
\right.
$$
\begin{equation}
+\frac{1}{t}(8.368 c^2 \ln \frac{t}{c^2}-3.408 c^2)+
 \frac{1}{t^2}(9.126 c^4
\ln \frac{t}{c^2}+2.26 c^4) +
\label{402}
\end{equation}
$$
\left. +\frac{1}{t^3}(4.043 c^6 \ln \frac{t}{c^2} +7.41 c^6)
 + ... \right\}
$$
and for $\delta\phi(t)$ we use the leading approximation calculated in ref.
\cite{23} and \cite{18a}-\cite{24b},
\begin{equation}
\delta\phi(t) =
\frac{3-2s^2}{2s^2c^2}\left
\{-\frac{\pi^2}{3}(\frac{\hat{\alpha}_s(m_t)}{\pi})
t +\frac{1}{16s^2c^2}(\frac{\bar{\alpha}}{\pi})t^2
\tau_b^{(2)}(\frac{h}{t})\right\} \;\; ,
\label{403}
\end{equation}
where the function $\tau_b^{(2)}$ is given by the table \ref{tab1}
 \cite{24a}-\cite{24b} for $m_H / m_t <4$.
 For $m_H / m_t >4$ we use the expansion \cite{18a}-\cite{18b}:
$$
\tau_b^{(2)}(\frac{h}{t}) = \frac{1}{144}[311+24\pi^2+282\ln r+90\ln^2 r
-4r(40+6\pi^2+15\ln r+18\ln^2 r) +
$$
\begin{equation}
+\frac{3}{100}r^2(24209-6000\pi^2-45420\ln r-18000\ln^2 r)],
\label{403a}
\end{equation}
where $r=t/h$.
  For $m_t = 175$ GeV and
 $m_H = 300 $ GeV
$$ \tau_b^{(2)} = 1.245\;\; .
$$

\underline{Asymmetries}
are calculated by using eqs. (\ref{18}) -- (\ref{22})
with the loop
corrected values of $g_A$ and $g_V$.

\subsection{Auxiliary functions $F_t$ and $F_h$}

The functions $F_t(t)$ and $F_h(h)$ are used in eqs. (\ref{34}) --
(\ref{36}).

\begin{equation}
F_t(t) = \left\{ \begin{array}{ccc}
2(1-\sqrt{4t-1} \arcsin \frac{1}{\sqrt{4t}}) & , & 4t > 1 \\
 & & \\
2(1-\sqrt{1-4t} \ln \frac{1+\sqrt{1-4t}}{\sqrt{4t}}) & , & 4t < 1
\end{array}
\right.
\label{irok1}
\end{equation}

\begin{equation}
F_h(h) = \left\{ \begin{array}{ccc}
1+(\frac{h}{h-1}-\frac{h}{2})\ln~ h + h\sqrt{1-\frac{4}{h}}
\ln(\sqrt{\frac{h}{4}-1} +\sqrt{\frac{h}{4}}) & , & h > 4 \\
 & & \\
1+(\frac{h}{h-1}-\frac{h}{2})\ln~ h - h\sqrt{\frac{4}{h}-1}
\arctan \sqrt{\frac{4}{h}-1} & , & h < 4
\end{array}
\right.
\label{irok2}
\end{equation}

\begin{equation}
F'_h(h) = \left\{ \begin{array}{ccc}
-1+\frac{h-1}{2}\ln~ h +(3-h)\sqrt{\frac{h}{h-4}}\ln(\sqrt{\frac{h}{4}-1}
+\sqrt{\frac{h}{4}}) & , & h > 4  \\
 & & \\
-1+\frac{h-1}{2}\ln~ h
+(3-h)\sqrt{\frac{h}{4-h}}\arctan(\sqrt{\frac{4-h}{h}})& , & h < 4
\end{array}
\right.
\label{irok3}
\end{equation}

\newpage
\begin{table}[h]\centering
\caption[]
{ Functions $A(\frac{m_H}{m_t})$ and $\tau^{(2)}(\frac{m_H}{m_t})$
from \cite{24a}-\cite{24b}.
  }
\label{tab1}
\vspace{0.5mm}
\begin{tabular}{|r|r|r|}
\hline
 $\frac{m_H}{m_t}$ & $A(\frac{m_H}{m_t})$ & $\tau^{(2)}(\frac{m_H}{m_t})$
 \\ \hline
  .00  &    .739  & 5.710 \\
  .10  &   1.821  & 4.671 \\
  .20  &   2.704  & 3.901 \\
  .30  &   3.462  & 3.304 \\
  .40  &   4.127  & 2.834 \\
  .50  &   4.720  & 2.461 \\
  .60  &   5.254  & 2.163 \\
  .70  &   5.737  & 1.924 \\
  .80  &   6.179  & 1.735 \\
  .90  &   6.583  & 1.586 \\
 1.00  &   6.956  & 1.470 \\
 1.10  &   7.299  & 1.382 \\
 1.20  &   7.617  & 1.317 \\
 1.30  &   7.912  & 1.272 \\
 1.40  &   8.186  & 1.245 \\
 1.50  &   8.441  & 1.232 \\
 1.60  &   8.679  & 1.232 \\
 1.70  &   8.902  & 1.243 \\
 1.80  &   9.109  & 1.264 \\
 1.90  &   9.303  & 1.293 \\
 2.00  &   9.485  & 1.330 \\
 2.10  &   9.655  & 1.373 \\
 2.20  &   9.815  & 1.421 \\
 2.30  &   9.964  & 1.475 \\
 2.40  &  10.104  & 1.533 \\
 2.50  &  10.235  & 1.595 \\
 2.60  &  10.358  & 1.661 \\
 2.70  &  10.473  & 1.730 \\
 2.80  &  10.581  & 1.801 \\
 2.90  &  10.683  & 1.875 \\
 3.00  &  10.777  & 1.951 \\
 3.10  &  10.866  & 2.029 \\
 3.20  &  10.949  & 2.109 \\
 3.30  &  11.026  & 2.190 \\
 3.40  &  11.098  & 2.272 \\
 3.50  &  11.165  & 2.356 \\
 3.60  &  11.228  & 2.441 \\
 3.70  &  11.286  & 2.526 \\
 3.80  &  11.340  & 2.613 \\
 3.90  &  11.390  & 2.700 \\
 4.00  &  11.436  & 2.788 \\
\hline
\end{tabular}
\end{table}

\newpage
%\large
%\begin{center}
%{\bf Part II}
%
%\vspace{3mm}

\chapter{LEPTOP THEORETICAL UNCERTAINTIES}

\normalsize
\section{General remarks}

With the increasing experimental accuracy of
the precision measurements at LEP
and SLC the question of the accuracy of
the theoretical  predictions  becomes one of
great importance. There are several
major sources of uncertainties:

1. The degree of accuracy with which
any given observable is extracted from the
experimental data. This involves theoretical models and computer simulations
(Monte Carlo etc.) which deal with purely electromagnetic and strong
interactions, for instance, in extracting $m_W$ from the
$p\bar{p}$-collider data, or $Z$-observables (in particular for separate
quark channels) from $e^+e^-$-data.  We call them "extraction"
uncertainties.

2. Uncertainties in the input parameters ($\bar{\alpha}$, $m_Z$,
$G_{\mu}$, $m_b$, $m_{\tau}$ etc.),
so  called "parametric" uncertaities.

3. The degree of accuracy with which the theoretical expression for a given
electroweak observable is derived within the MSM or  some of its
generalizations ("theoretical" uncertainties).

In this paper we shall mainly deal with the third kind of uncertainties
(in the framework of the MSM),
although for some of observables, such as $R_b$ or $A^b_{FB}$ or $Q_{FB}$,
they may  not be fully separable from
those of the first kind. One also has to keep in
mind that some of the physical quantities we shall discuss such as
$\Gamma_h, \Gamma_e, g_A$, are,  strictly
speaking, not primary, but derived observables.

There are several types of observables with specific uncertainties for
each type:

1. "hadron-free" observables: $m_W\;\; (m_W/m_Z)$, $\Gamma_l$, $g_A$,
$g_V/g_A$.

2. Observables of hadronic decays of $Z$ boson.

3. Observables of the low-energy weak processes: $\nu e$-scattering, $\nu
N$-scattering, atomic parity violation parameters.

Let us finish this introductory section with the estimates of the
"parametric" uncertainties for the $Z$-decay parameters and
the $W$ boson mass. Among the three basic input parameters $\bar{\alpha}\;,
G_{\mu}$ and $m_Z$ the largest uncertainty comes from $\bar{\alpha} \equiv
\alpha(m_Z): \bar{\alpha}^{-1} = 128.87(12)$. The relative uncertainty in
$m^2_Z$ is an order of magnitude smaller $(m_Z = 91.1887(44))$; that in
$G_{\mu}$ is 50 times smaller $(G_{\mu} = 1.16639(2) \cdot 10^{-5}
\mbox{\rm GeV}^{-2})$.
 Induced by the uncertainty in $\bar{\alpha}$ the absolute
uncertainty in $m_W/m_Z$ is close to $ 2\cdot 10^{-4}$, that in $g_V/g_A$ is
close to $1\cdot 10^{-3}$, in $\Gamma_h$ it equals  0.8 MeV and in
$\sigma_h$ it is 0.001 nb. Finally, in $\Gamma_b$ it is 0.15 MeV.
All these uncertainties come mainly from the uncertainty of the
$\bar{\alpha}$-Born approximation due to that in the parameter $s^2 \equiv
\sin^2\theta = 0.2312(3)$.

Let us stress that relative parametric uncertainty in $g_V/g_A$ is quite
large:  about 1.5\%.  The experimental accuracy with which $g_V/g_A$ is
measured is becoming close to the uncertainty due to $\delta\bar{\alpha}$.
Improvement of the accuracy of $\bar{\alpha}$ calls for new measurement of
$e^+e^- \to hadrons$ cross-section below $J/\psi$ resonanse. Measurement of
$(g-2)_{\mu}$ with high accuracy will also help.

\section{Uncertainties in $V_i$}

Theoretical uncertainties come from as yet not calculated Feynman diagrams
or not calculated terms in a given diagram. The smallest calculated terms in
the $W$ and $Z$  selfenergies are
the virtual higgs corrections of the order
$\alpha^2_W t^2$ \cite{18a}-\cite{24b}, and the two gluon correction
$\alpha_W\hat{\alpha}^2_s t$ \cite{17} to the top quark loop. These
corrections produce universal shifts $\delta V_i$ in the functions
$V_i$ which describe radiative corrections in the approach developed
in \cite{2}:  \begin{equation} \delta V^{t^2}_i =
-\frac{\bar{\alpha}}{\pi} \frac{A(m_h/m_t)t^2}{16 s^2c^2} \label{101}
\end{equation}
\begin{equation}
\delta V^{\alpha^2_s}_i = -1.2 \hat{\alpha}^2_s(m_t)t
\label{102}
\end{equation}
Here  and in what follows we denote by $\delta$ the corrections, while by
$\Delta$ -- the uncertainties.

According to \cite{200}, still not
completely  calculated correction to (\ref{101}), of the order of
$\bar{\alpha}t$, can be numerically close to (\ref{101}).
The same, according to the literature, is valid for uncertainties in eq.
(\ref{102}). In order to have correct asymptotic behaviour of the
uncertainties at $t\gg 1$ we assume that uncertainties correslponding to
expressions (\ref{101}) and (\ref{102}) are derived by multiplying both of
them by $2/t$.
 For $m_H = 300$ GeV $,\;\; m_t = 175$ GeV
and $ \hat{\alpha}_s(m_Z) = 0.125$
our estimates are  $\Delta V_i^{t^2} = \pm 0.06\;,
\Delta V_i^{\alpha^2_s} = \pm 0.03$; corresponding
uncertainties in the observables are
presented in the Table 1.

\section{Uncertainties in hadronic $Z$ decays}

In hadronic decays we separate decays into light quarks $q\bar{q}$ (where $q
= u,d,s,c)$ and the decay into $b\bar{b}$. Concerning $\Gamma_q$ the next
not yet calculated Feynman diagrams are the gluon corrections to the $Zqq$
electroweak vertex "triangle" involving $W$ or $Z$. The electroweak
corrections to the vertices in the $Z \to qq$ decays are given by the
following equation:
\begin{equation}
\delta \Gamma_q = 24sc\Gamma_0
2T^q_3\{(1-4|Q_f|s^2) F_{Vq} + F_{Aq}\}\;\;.
\label{103}
\end{equation}

 Substituting numbers in eq. (\ref{103}) we get:
\begin{equation}
\Delta\Gamma_u = \Delta\Gamma_c = -1.9 \; \mbox{\rm MeV}\;\;,
\label{104}
\end{equation}
\begin{equation}
\Delta\Gamma_d = \Delta\Gamma_s = -2.0 \; \mbox{\rm MeV}\;\;.
\label{105}
\end{equation}

Taking into account 4 light quark flavors and  multiplying the results
(\ref{104}) and (\ref{105}) by $\hat{\alpha}_s(m_Z)/\pi$ we get the following
estimate of the uncertainty in $Z$ width into light quarks due to the
gluonic perturbation of $Zqq$ vertex:
\begin{equation}
\sum^4_1 \Delta\Gamma_q^{\bar{\alpha}
\hat{\alpha}_s} = \pm 0.3 \; \mbox{\rm MeV}
\label{106}
\end{equation}

For the $Z \to b\bar{b}$ decay the leading term $\sim
\alpha_W\hat{\alpha}_s t$
\cite{23},
and the potentially  next to the leading term
$\sim \alpha_W \hat{\alpha}_s\ln t$
\cite{201} come from virtual gluons in the
$ttW$ vertex triangle. Calculated in \cite{23} the leading $\alpha_s$
correction to electroweak vertex lead to the following
 correction to $\Gamma_b$:
\begin{equation}
\Delta
\Gamma_b^{\bar{\alpha}\hat{\alpha}_s t} = -\frac{\bar{\alpha}}{\pi} \Gamma_0
\frac{(3 - 2s^2)}{2s^2c^2} (-\frac{\pi^2}{3})
\frac{\hat{\alpha}_s(m_t)}{\pi} t = 0.7 \; \mbox{\rm MeV}\;\;,
\label{107}
\end{equation}
where for numerical estimate we substitute $m_t$ = 175 GeV.

Calculated in \cite{201} $\hat{\alpha}_s$ correction to $\ln t$ term appears
to be much smaller:
\begin{equation}
\delta \Gamma_b^{\bar{\alpha}\hat{\alpha}_s\ln t} =
-\frac{\bar{\alpha}}{\pi} \Gamma_0 \frac{(3 - 2 s^2)^2}{16 s^2c^2}
\frac{7}{81} \frac{\hat{\alpha}_s(m_t)}{\pi} \ln t \simeq 0.002 \;
\mbox{\rm MeV}
\label{108}
\end{equation}

If we suppose that uncalculated terms are of the order of the last
calculated one we can safely neglect this type of uncertainties.

Uncertainty in $m_b$ also shows itself in $\Gamma_{b}$
value.  However, it is comparatively small. Even if we assume that $m_b$ is
known with accuracy $\pm 300$ MeV,
 the shift in $\Gamma_b$ is $\pm 0.17$ MeV.
There is a specific to $Z \to b\bar{b}$ decay virtual higgs gaugeless
correction to the vertex $\delta\phi(t,h)$ consisting of terms of the order
$\alpha_W^2 t^2\;, \alpha^2_W t \ln t\;, \alpha_W^2 t$ and so on.
Only the first of them have been calculated \cite{18a}-\cite{24b}
 :
\begin{equation}
\delta \phi^{t^2} = \frac{3-2s^2}{2s^2c^2} \frac{\bar{\alpha} t^2}{16\pi
s^2c^2} \tau^{(2)}_b\;\;.
\label{109}
\end{equation}
If we consider this
leading term as an upper bound of the "higgs theoretical uncertainty" in
$\Delta \phi$ than we get:
\begin{equation}
\Delta \Gamma^{t^2}_b
= -\frac{\bar{\alpha}}{\pi} \Gamma_0 \Delta\phi^{t^2} =
-\frac{\bar{\alpha}}{\pi}\Gamma_0 \frac{3-2s^2}{2s^2c^2} \frac{\bar{\alpha}
 t^2}{16\pi s^2c^2} \tau^{(2)}_b = 0.02 \; \mbox{\rm MeV}\;\;,
 \label{110}
 \end{equation}
where we substitute $m_t = 175$ GeV $,\;\; m_H = 300$ GeV.
This uncertainty in $\Gamma_b$ is much smaller than those considered above
and could be neglected. However, correction of the order of $\hat{\alpha}^2_s
 (m_t ) t$ to
$\phi (t), \delta \phi^{\alpha^2_s}$, may lead to larger correction to
 $\Gamma_b$.

To estimate its value let us suppose that it's ratio to the calculated
 correction
$\sim \hat{\alpha}_s t$ is equal to the ratio of the calculated corrections
$\sim \hat{\alpha}^2_s t$ and $\sim \hat{\alpha}_s t$ to functions $V_i$:

\begin{equation}
\frac{\delta \phi ^{\alpha ^2_s}}{\delta \phi ^{\alpha _s}} =
\frac{\delta _3 V_i}
     {\frac{\hat{\alpha}_s (m_t)}
     {\pi}(-2.86) t} ,
\label{11111}
\end{equation}
\begin{equation}
\delta \phi ^{\alpha ^2_s} = - \frac {3-2s^2}{2s^2 c^2} \frac{1.37 \pi}{3}
\hat{\alpha}^2_s (m_t ) t .
\label{11112}
\end{equation}
Taking this estimate as a source of uncertainty
 in $\Gamma_b$, for $m_t$ = 175
 GeV we get:
$\Delta \Gamma ^{\alpha ^2_s}_b$ = 0.1 MeV,
 which dominates over (\ref{108}) and
 (\ref{110}).

\section{Procedure to estimate total theoretical accuracy of LEPTOP}

In order to estimate the total theoretical uncertainties for the observables
we implement in LEPTOP the procedure proposed by D.Bardin and G.Passarino
\cite{202}. We choose several options to the preferred formulas used in
LEPTOP and make variations of these formulas. Usually each option consists
in the addition of an extra term corresponding to a rough guess of the value
of the uncalculated higher order terms. We then make all possible
combinations of these options, i.e. $2^n$ in total, where $n$ is the number
of options.  Among all these $2^n$ combinations we  locate those yielding
the minimum and the maximum values of the observables and took as the
estimate of the theoretical errors their deviations from the central values.

We consider the following options:
\begin{DLtt}{123456789012}
\item[Options 1,2] Correction $\delta V_i^{t^2}$
                   given by eq. (\ref{101})  multiplied by $2/t$
                   is added to (option 1)
                   (or subtracted from (option 2))
                   the three functions $V_A\;, V_m$ and $V_R$.
\item[Options 3,4] The same as in options 1,2 with
                   $\delta V_i^{\alpha^2_s}$ from
                   eq. (\ref{102}).
\item[Options 5,6] The expression for $\Gamma_q$
                   is modified by adding (option 5)
                   or subtracting (option 6)
                    $\sum^4_1\Delta\Gamma_q = 0.3$ MeV
                   in order to take into
                   account gluon corrections to electroweak $Zqq$
                   triangle vertices.
\item[Option 7] Correction $\delta\phi^{\alpha^2_s}$
                   eq. (\ref{11112}) is added to  function $\phi (t)$.
\end{DLtt}

\newpage

\begin{table}[h]\centering
\caption[]
{\it The effect of the working options of {\tt LEPTOP}
 on theoretical errors.
The first two lines indicate parametric uncertainties caused by
 $\delta s^2 =
0.0003$ (which is equivalent to $\delta\bar{\alpha}^{-1} = 0.12$)
 and by
$\delta m_b = 0.3$GeV. The next six lines refer to the intrinsic
 theoretical
uncertainties. This table was calculated with
 $m_t$=175 GeV, $M_H$=300 GeV
and $\hat{\alpha_s}$=0.125.}
\label{ta18}
\vspace{2.5mm}
\begin{tabular}{|c|r|r|r|r|r|r|r|r|r|r|}
\hline
   & $m_W$ & $\Gamma_{l}$ & $\sin^2\theta^l_{eff}$ & $\sigma^h_0(nb)$ &
  $ \Gamma_Z$ & $\Gamma_h$ & $R_l$ & $\Gamma_b$ & $R_b\times$ &
  $R_c\times$ \\
  & MeV & MeV & & & MeV & MeV & &MeV & $10^5$ &
   $10^5$\\ \hline
 $\delta s^2$ &
 16    & .015    & .00031 & .0014 &  .8 &  .8 &  .0055 & .15 &
   1.1 & 1.8
 \\ \hline
  $\delta m_b$ &
  -    & -       & -      & .0023 &  .2 &  .2 &  .0029 & .24 &
   11.0 & 2.4
 \\ \hline \hline
  $\Delta V_i^{t^2}$ &
  9    & .015    & .00005 & .0002 &  .5 &  .4 &  .0009 & .08 &
     .2 &  .3
 \\ \hline
 $\Delta V_i^{\alpha^2_s}$ &
  5    & .008    & .00003 & .0001 &  .3 &  .2 &  .0005 & .04 &
     .1 &  .2
 \\  \hline
 $\Delta \Gamma_q$ &
  -    & -       & -      & .0029 &  .3 &  .3 &  .0037 & -   &
    3.8 & 1.5
 \\ \hline
 $\Delta \phi^{\alpha^2_s}$ &
  -    & -       & -      & .0010 &  .1 &  .1 &  .0012 & .10 &
    4.6 & 1.0
 \\ \hline
total &+13 &+.023 &+.00008 &+.0032 &+1.2 &+1.0 &+.0063 &+.23 &+8.7&
+1.9 \\
 &$-13$ &$-.023$ &$-.00008$ &$-.0042$ &$-1.1$ &$-.9$
 &$-.0051$ &$-.13$ &$-4.1$&
$-2.9$ \\ \hline
\end{tabular}
\end{table}
\normalsize

%\newpage
%\begin{center}
%\large{\bf Part III\\

%\vspace{3mm}
%
%LEPTOP CODE}
%\end{center}

\chapter{ LEPTOP  program}
\section{Introduction}

LEPTOP code allows the calculations of
LEP-I electroweak observables in the
framework of the Minimal
Standard Model and fitting the experimental data for determination
of $m_t$, $\hm$ and $\hat \alpha_s(M_Z)$.
LEPTOP package has been written in the Fortran language and installed on
IBM VM and HP-SUN UNIX operating systems. It uses the subroutines from the
CERN program library \cite{cernlib}.
The code is maintained with the CMZ \cite{cmz} source code management
system.
The Fortran code is available
either in the form of a compilable source file $leptop.f$
 or in the CMZ ASCII readable file $leptop.car$. In the following
sections we describe  simple useage of the LEPTOP package.
The user interface part of LEPTOP was written in the object oriented
style, which allows a simple use of it without deep knowledge of the
Fortran language. The data encapsulation principle and hiding the
internal variables are applied. This significantly reduces the amount
 of information the user should care about improves the robustness of the
program.

 In order to obtain the
maximum accuracy we used whenever possible double-precision variables
in the internal calculations.
However the user interface was written in a single precision in order to
simplify the use of the program, make easy interface with other
standard packages and avoid user's bugs typical for the mixture
of single and double precision variables. The 7 digit precision
(for 32 bit machines) in the
input and output is adequate for the present experimental and
theoretical errors. The units used in the program are GeV and nanobarns,
unless explicitly stated.

This document is a {\bf Reference manual} and {\bf User's Guide}.
 Whener possible
we tried to follow $cernman$ style and notations \cite{hbook}.
An example of a simple program is given in the section 3.8.

This report, including
the writeup and Fortran code, is available under WWW \cite{www} on
\Lit{http://cppm.in2p3.fr/leptop/intro_leptop.html}.

\section{Initialisation}

\Shubr{LTINIT}{(ISTAT)}

\Action
The setting of the default values of all variables is done.
\Lit{LTINIT(0)} should be called before any other \Lit{LEPTOP} routine.

\begin{DLtt}{123456}
\item[{\rm\bf Input parameters:}]
\item[ISTAT]    Level of the initialisation.
\begin{DLtt}{12345678}
\item[ISTAT=0]    means complete
                  initialisation from scratch.
\item[ISTAT=1]    means recalculating
                  constants after the eventual update of the data.
                  This option is needed
                  only for advanced users, because in the
                  user interface it is done automatically.
\end{DLtt}
\end{DLtt}

\section{Input and output}

The output of the program is directed into the unit 6.
The unit 19 is reserved for the internal scratch output.
The amount of output is controlled via \Lit{PRNT} flags
 (see \Lit{LTFLAG} routine).

\section{Setting of flags and options}

\Shubr{LTFLAG}{(KEY,ID)}

\Action
Setting on and off the flags according to the character
variable \Lit{KEY}, identifying the group of flags,
 and integer index \Lit{ID}, identifying specific flag of that group.

 Positive values of \Lit{ID}
sets the corresponding flag on, negative value of the \Lit{ID} sets
it off. Zero value of the ID sets off all the flags of the corresponding
group. By default all the flags and options are switched off.

\begin{DLtt}{12345}
\item[{\rm\bf Input parameters:}]
\item[KEY] Character variable identifying the group of flags.
\begin{DLtt}{1234567}
\item['PRNT'] Group of flags controlling the output level.
\item['OPT'] Group of options modifying the prefered formulas of LEPTOP.
             See section 2.4 for details.
\end{DLtt}
\item[ID] Numerical identifier of the flag within the group.
\item[ ] The meaning of different flags is given below:
\begin{DLtt}{1234567}
\item['PRNT'] Allowed values for \Lit{ID} are:
\begin{DLtt}{123}
\item[1] Low level printout for debug purpose in the process of calculation.
\item[2] Detailed printout during the fit. It may produce very big outputs.
\item[3] Detailed printout after the fit. Printout of the table with MSM
         and Born predictions for some observables in the style of the
         table in the paper \cite{9}.
\item[4] Printout during the initialisation of several default constants
         and experimental inputs.
\item[8] Printout during the setting of the variables and
         experimental data is done by the
         user interface routines \Lit{LTPUT, LTFPUT, LTFCOR, LTFUSE} etc.
\item[9] Printout of the calculated observables, setting constants and
         experimental data is done by the user interface routines
         \Lit{LTGET, LTFGET, LTFIT1, LTFIT2} etc.
\end{DLtt}
\item['OPT'] Allowed values for \Lit{ID} are:
\begin{DLtt}{123}
\item[1]  Option 1: functions $V_A, V_M, V_R$ are increased by
          the terms $(2/t) \delta V_{i}^{t^2}$.
\item[2]  Option 2: same variation as option 1, but with negative sign.

\item[3]  Option 3: functions $V_A, V_M, V_R$ are increased by
          the terms $(2/t) \delta V_{i}^{\alpha^{2}_{s}}$.
\item[4]  Option 4: same variation as option 3, but with negative sign.
\item[5]  Option 5: the quark widths of Z are increased in order to take
          into account estimated QCD corrections to
          Zqq triangle vertices.
\item[6]  Option 6: same variation as option 5, but with negative sign.

\item[7]  Option 7: function $\phi(t)$ is increased by the term
          $\delta \phi^{\alpha^2_s}$
\item[8]  Option 8: the b-quark width of Z is increased by 7 MeV.
\end{DLtt}
\item['MNUNIT'] Changing the Fortran output unit of MINUIT to unit \Lit{ID}.
                The default is the
                the output on the scratch file on unit 19. Allowed values for
                \Lit{ID} are from 1 to 99. To restore the output on the
                scratch file one should assign \Lit{ID} to \Lit{-19}.
\item['MNSAVE'] Changing the SAVE output unit of MINUIT to unit \Lit{ID}.
                The default unit is 7.
\item['MNREAD'] Changing the Fortran input unit of MINUIT to unit \Lit{ID}.
                The default value of \Lit{ID} is \Lit{-5}, which means
                that MINUIT is working in Fortran-callable mode.
                {\bf Attention:} using this flag switch MINUIT to the
                 data-driven mode, when user should provide on unit
                \Lit{ID} the complete set of MINUIT commands.
                If needed, the user can provide his own version of
                the function
                FCNLB.
\item['MNPRNT'] sets the print level of MINUIT.
                The default value of \Lit{ID} is \Lit{-1}, which means
                no output of MINUIT except from MINUIT \Lit{SHOW} command.
                Other possibles values of \Lit{ID} are in the range from 0
                to 3 according to MINUIT \Lit{SET PRIntout} command.
\item['MNEPS']  informs MINUIT that the relative floating point arithmetic
                precision is $10^{\Lit{ID}}$.
                The default precision is $10^{-10}$.
\end{DLtt}
\end{DLtt}
\Remark
\Lit{LTFLAG} is the only routine that can be called before
\Lit{LTINIT(0)}. This allows to suppress the printout during the
initialisation stage.

\Examples
\begin{XMP}
 CALL LTFLAG('OPT',7)
\end{XMP}
      This call sets on the option 7.
\begin{XMP}
CALL LTFLAG('PRNT',-4)
\end{XMP}
      This call sets off the printout of the fit results, if it
      has been set on previously. \\ \\

\section{Setting the constants and experimental data}

\Shubr{LTPUT}{(NAME,DATA)}

\Action
Setting the constant defined by keyword \Lit{NAME} to the
value defined by the variable \Lit{DATA}.
\begin{DLtt}{12345}
\item[{\rm\bf Input parameters:}]
\item[NAME] Character variable containing the name of the constant to be
           modified.
\begin{DLtt}{1234567}
\item['MT'] Mass of the top quark $\tm$.
\item['MH'] Mass of the Higgs boson $\hm$.
\item['MZ'] Mass of the Z boson $\zm$.
\item['ALBAR'] Electromagnetic coupling constant $\bar \alpha ( \zm )$.
\item['ALSHAT'] Strong coupling constant $\alphahat_s ( \zm )$.
\item['GFERMI'] Weak Fermi coupling constant $G_{\mu}$
                determined from the muon
                lifetime measurements.
\end{DLtt}
\item[DATA] New value of the constant.
\end{DLtt}

\Remark
   If \Lit{PRNT} flag number 8 was set before the call
                  to \Lit{LTPUT} the new value will be printed.

\Example
\begin{XMP}
   CALL LTPUT('MH',300.)
\end{XMP}
   This call will set the Higgs mass to 300 GeV. \\ \\

\Shubr{LTFPUT}{(TYPE,NAME,DATA)}

\Action
Setting the values of experimental data, including observables and their
errors, for the use in the fit of experimental data with LEPTOP
formulas.
\begin{DLtt}{12345}
\item[{\rm\bf Input parameters:}]
\item[TYPE] Character variable selecting the type of data to be set.
\begin{DLtt}{123456789}
\item['VALUE'] The central value of the experimental data is set.
\item['ERROR'] The error on the experimental data is set.
\item['MARSEILLE'] Experimental data presented at Marseille-93
                   conference are loaded in case \Lit{NAME='ALL'}.
\item['MORIOND94'] Same, but for Moriond-94 conference.
\item['GLASGOW'] Same, but for Glasgow-94 conference.
\item['MORIOND95'] Same, but for Moriond-95 conference.
\end{DLtt}
\item[NAME] Character variable identifying the experimental observable.
            The following observables may be used in the fit:
\begin{DLtt}{12345678}
\item['GZ'] Total width of Z boson $\gz$.
\item['SIGH'] The hadronic pole cross section
              $\s0h = (12 \pi/\zm^2)(\ge \gh/\gz^2)$.
\item['RL'] Ratio $R_l = \gh / \gl$ of hadronic to leptonic
            partial width of Z.
\item['AFBL'] Forward-backward lepton asymmetry at $\z0$ pole.
\item['RB'] Ratio $R_b = \gb / \gh$.
\item['MWMZ'] Ratio $\wm / \zm$ of W-boson to Z-boson masses.
\item['ATAU'] $A_{\tau}$ from $\tau$ polarization.
\item['AETAU'] $A_{e}$ from $\tau$ polarization asymmetry.
\item['AFBB'] Forward-backward b-quark asymmetry at $\z0$ pole.
\item['AFBC'] Forward-backward c-quark asymmetry at $\z0$ pole.
\item['QFB'] Measurement of forward-backward quark asymmetry
             translated in a value for $\stes_{eff}$.
\item['S2NUN'] Measurement of the ratio of neutral to charged current
               cross-sections translated into $\stes_W$.
\item['ALR'] Left-right polarization asymmetry $A_{LR}$, measured
             at SLAC by SLD.
\item['MT'] Top quark mass $\tm$ measured by CDF and D0 at Fermilab.
\item['RC'] Ratio $R_c = \gc / \gh$ of charm to hadron partial widths.
\item['ALL'] used to set all observables from the data set defined by
             the keyword \Lit{TYPE}.
\end{DLtt}
\item[DATA] Numerical value of the experimental data.
\end{DLtt}
\Remark
By default Glasgow-94 experimental data are loaded during the
initialisation with \Lit{LTINIT(0)}.
\Examples
\begin{XMP}
   CALL LTFPUT('VALUE','GZ',2.4974)
\end{XMP}
  This call sets the experimental value of $\gz$ to 2.4974 GeV.
\begin{XMP}
   CALL LTFPUT('ERROR','SIGH',0.12)
\end{XMP}
  This call sets the experimental value of the error on the peak
hadron cross section to 0.12 nb.
\begin{XMP}
   CALL LTFPUT('GLASGOW','ALL',dummy)
\end{XMP}
  This call sets the input of all experimental data as presented at
 Glasgow-94 conference. \\ \\

\Shubr{LTFCOR}{(NAME1,NAME2,DATA)}

\Action
Set the values of correlation coefficients on the experimental
errors,
for the use in the fit of experimental data with LEPTOP
formulas.
\begin{DLtt}{12345}
\item[{\rm\bf Input parameters:}]
\item[NAME1] Character variable identifying the first
             experimental observable, as defined
             in the \Lit{LTFPUT} description.
\item[NAME2] Character variable identifying the second
             experimental observable, as defined
             in the \Lit{LTFPUT} description.
\item[DATA] Numerical value of the correlation coefficient between the two
            experimental data.
\end{DLtt}
\Example
\begin{XMP}
   CALL LTFCOR('GZ','SIGH',-0.110)
\end{XMP}
  This call set the correlation coefficient between the experimental errors
  of $\gz$ and $\s0h$ equal to $\rho_{\gz \s0h} = -0.11$. \\ \\

\section{Access to the constants, observables and experimental data}

\Shubr{LTGET}{(NAME,DATA*)}

\Action
Extracting the constant or the theoretical prediction for the
observable  defined by keyword \Lit{NAME} to the
variable \Lit{DATA}.
\begin{DLtt}{12345}
\item[{\rm\bf Input parameters:}]
\item[NAME] Character variable containing the name of the constant
or theoretical value of the observable to be extracted.
\begin{DLtt}{1234567}
\item['MT'] Mass of the top quark $\tm$.
\item['MH'] Mass of the Higgs boson $\hm$.
\item['MZ'] Mass of the Z boson $\zm$.
\item['ALBAR'] Electromagnetic coupling constant $\bar \alpha ( \zm )$.
\item['ALSHAT'] Strong coupling constant $\alphahat_s ( \zm )$.
\item['GFERMI'] Weak Fermi coupling constant $G_{\mu}$ defined by muon
                lifetime measurements.
\item['GV'] Vector coupling of the electron.
\item['GA'] Vector axial coupling of the electron.
\item['SIN2E'] The effective electro-weak mixing angle $\stes_{eff}$
               for leptons.
\item['SIN2B'] The effective electro-weak mixing angle $\stes_{eff}$
               for b quarks.
\item['MW'] Mass of the W boson $\wm$.
\item['GNU'] Partial width of Z boson into $\nu \bar \nu$.
\item['GE'] Partial width $\ge$ of $\z0$ into $e^+e^-$.
\item['GMUON'] Partial width $\gm$ of $\z0$ into $\mu^+\mu^-$.
\item['GTAU'] Partial width $\gt$ of $\z0$ into $\tau^+\tau^-$.
\item['GU'] Partial width $\gu$ of Z into $u \bar u$ quarks.
\item['GD'] Partial width $\gd$ of Z into $d \bar d$ quarks.
\item['GS'] Partial width $\gs$ of Z into $s \bar s$ quarks.
\item['GC'] Partial width $\gc$ of Z into $c \bar c$ quarks.
\item['GB'] Partial width $\gb$ of Z into $b \bar b$ quarks.
\item['GINV'] Invisible width of Z boson $\Gamma_{inv}$.
\item['GH'] Hadron width of Z boson $\gh$.
\item['GZ'] Total width of Z boson $\gz$.
\item['RL'] Ratio $R_l = \gh / \gl$ of hadronic to leptonic
            partial width of Z.
\item['RC'] Ratio $R_c = \gc / \gh$ of charm to hadron partial widths.
\item['RB'] Ratio $R_b = \gb / \gh$.
\item['AFBL'] Forward backward lepton asymmetry $\afb^l$.
\item['AFBC'] Forward-backward c-quark asymmetry at Z pole.
\item['AFBB'] Forward-backward b-quark asymmetry at Z pole.
\item['ALR'] Left right asymmetry $\alr$.
\item['SIGH'] The hadronic pole cross section
              $\s0h = (12 \pi/\zm^2)(\ge \gh/\gz^2)$.
\end{DLtt}
\item[{\rm\bf Output parameter:}]
\item[DATA] numerical value of the constant or the result
            of the calculation for the prediction.
\end{DLtt}
\Remark
\begin{UL}
\item The extracted value may depend on the change of the constants by
         the previous call to \Lit{LTPUT}. So to get a consistent set
         of observables one should make all calls to \Lit{LTGET}
         after the last call of \Lit{LTPUT}.
\item   If \Lit{PRNT} flag number 9 was set before the call
                  to \Lit{LTGET} the value of the observable will be printed.
\end{UL}
\Example
\begin{XMP}
   CALL LTGET('MW',AMW)
\end{XMP}
   After this call the variable AMW will contain
   the W boson mass in GeV.  \\ \\

\Shubr{LTFGET}{(TYPE,NAME,DATA*)}

\Action
Extracting the values of experimental data, including observables and their
errors. It also can extract the values of the fitted parameters
 and their errors after the fit.
\begin{DLtt}{12345}
\item[{\rm\bf Input parameters:}]
\item[TYPE] Character variable selecting the type of data to be extracted.
\begin{DLtt}{12345678901}
\item['VALUE'] The central value of the experimental data is given.
\item['ERROR'] The error on the experimental data is given.
\item['FIT\_RESULT'] The central value of the fitted parameter after the
                    last call to \Lit{LTFIT1} or \Lit{LTFIT2}.
\item['FIT\_ERROR'] The corresponding parabolic error of the fitted parameter.
\item['FIT\_ERR+'] The corresponding positive error of the fitted
                  parameter.
\item['FIT\_ERR-'] The corresponding negative error.
\item['FIT\_GLB'] The corresponding global correlation coefficient
                 with other fitted parameter.
\end{DLtt}
\item[NAME] Character variable identifying the experimental observable
            as defined in the description of \Lit{LTFPUT} for
            \Lit{TYPE='VALUE'} or \Lit{TYPE='ERROR'}.
            If \Lit{NAME='ALL'} and \Lit{PRNT} flag 9 is set
            on, the table of all experimental data is printed.
            If \Lit{TYPE='FIT\_*'} the allowed values of \Lit{NAME}
            are: \Lit{MT}, \Lit{MH} and \Lit{ALSHAT}, corresponding to
            the fitted parameters.
\item[{\rm\bf Output parameters:}]
\item[DATA] Numerical value of the experimental data or fitted parameter.
\end{DLtt}
\Remark
\begin{UL}
\item   If \Lit{PRNT} flag number 9 was set before the call
                  to \Lit{LTFGET} the value of the observable will be printed.
\end{UL}
\Examples
\begin{XMP}
   CALL LTFGET('VALUE','MWMZ',RMWMZ)
\end{XMP}
  After this call the variable RMWMZ will contain the experimental ratio of
 W boson to Z boson masses $\wm / \zm$.
\begin{XMP}
   CALL LTFGET('ERROR','SIGH',ESIGH)
\end{XMP}
  After this call the variable ESIGH will contain the experimental value
  of the error on the peak hadron cross section $\s0h$ in nb.
\begin{XMP}
  CALL LTFLAG('PRNT',9) \\
  CALL LTFGET(' ','ALL',DUMMY)
\end{XMP}
  After these calls the table with all experimental data
  is printed.
\begin{XMP}
  CALL LTFIT1('MH',amh,emh,chi2)
  CALL LTFGET('FIT_ERR+','MH',emhpos)
  CALL LTFGET('FIT_ERR-','MH',emhneg)
\end{XMP}
  After these calls the variables \Lit{emhpos} and \Lit{emhneg} will
  contain the asymmetric errors on Higgs mass, which might be
  significantly different from the parabolic error \Lit{emh}
  due to the non-gaussian distributions. \\ \\

\section{Fitting experimental data with LEPTOP formulas}
   The fitting in LEPTOP is done using MINUIT package \cite{minuit}.
The fitted parameter values correspond to the minimum of the $\chi^2$
function:
\begin{equation}
\chi^{2}\, =\, \sum_{i,j=1}^{n}
                   \frac{
\left( E_i -T_i \left( \tm , \hm , \alphahat_s \right) \right)
\left( E_j -T_j \left( \tm , \hm , \alphahat_s \right) \right)
}
{ S_{ij} }
\end{equation}
\begin{equation}
\frac{1}{S_{ij}}\, =\,
\left( V_{ij} \right)^{-1}
\end{equation}
\begin{equation}
V_{ij} =  \rho_{ij} \delta E_i \delta E_j
\end{equation}

where the following notations were used:
\begin{DLtt}{123456}
\item[n] Number of experimental observables used in the fit.
\item[$E_i$] Experimental observable, which usually comes from the
             compilation of results of all LEP detectors.
\item[$T_i$] Theoretical prediction of LEPTOP as a function of
             parameters $\tm$, $\hm$ and $\alphahat_s$.
\item[$V_{i,j}$] Covariance matrix of errors.
\item[$\delta E_i$] The experimental error on the observable i.
\item[$\rho_{ij}$] The correlation coefficient
                   between the experimental errors
                   of the observable i and observable j.
\end{DLtt}

The initial values of parameters are used from the settings done
by the routine \Lit{LTPUT}. So the user can check the sensitivity of
the fit to the initial values of the parameters. It is the user's
responsibility to provide reasonable values of the initial
parameters, as the minimisation procedure may diverge and move the
parameters to the unphysical region, where the program might be not
protected.

\Shubr{LTFUSE}{(COMMAND,NAME)}

\Action
Instructing LEPTOP to use or not to use the particular experimental
data in the fit.
By default all experimental data listed in \Lit{LTFPUT} description
are used, except \Lit{'MT'}.
\begin{DLtt}{12345678}
\item[{\rm\bf Input parameters:}]
\item[COMMAND] Character variable selecting the command.
\begin{DLtt}{12345678}
\item['USE'] Command to use this data in the fit.
\item['NOUSE'] Command to exclude this data from the fit.
\end{DLtt}
\item[NAME] Character variable identifying the experimental observable
            as defined in the description of  \Lit{LTFPUT}.
            If \Lit{NAME='ALL'}
            all experimental data are affected.
\end{DLtt}
\Examples
\begin{XMP}
   CALL LTFUSE('USE','MT')
\end{XMP}
  Add the experimental value of $\tm$ from CDF to the fit.
\begin{XMP}
   CALL LTFUSE('NOUSE','QFB')
\end{XMP}
  Do not use the experimental value of $Q_{FB}$ in the fit. \\ \\

\Shubr{LTFIT1}{(PARAM,PAR1*,EPAR1*,CHI2*)}

\Action

Performing one parameter fit of the experimental data.
The actual parameter is selected by the character constant \Lit{PARAM}
from $\tm$, $\hm$ and $\alphahat_s$.
\begin{DLtt}{12345678}
\item[{\rm\bf Input parameters:}]
\item[PARAM] Character variable selecting the parameters.
\begin{DLtt}{12345678}
\item['MT'] Select the parameter  $\tm$.
\item['MH'] Select the parameter  $\hm$.
\item['ALS'] Select the parameter $\alphahat_s$.
\end{DLtt}
\item[{\rm\bf Output parameters:}]
\item[PAR1] Value of the fitted parameter.
\item[EPAR1] Estimate of the error of the fitted parameter.
\item[CHI2] Value of the $\chi^2$ of the fit.
\end{DLtt}
\Example
\begin{XMP}
   CALL LTFIT1('MH',AMH,EMH,CHI2)
\end{XMP}
 After this call the variable AMH contains the fitted Higgs mass $\hm$,
  the variable EMH contains the parabolic estimate of the error
 of the Higgs boson mass
 and CHI2 the $\chi^2$ of the fit. \\ \\

\Shubr{LTFIT2}{(PARAM,PAR1*,PAR2*,EPAR1*,EPAR2*,R12*,CHI2*)}

\Action

Performing two parameter fit of the experimental data.
The actual parameters are selected by the character constant \Lit{PARAM}
from $\tm, \hm$ and $\alphahat_s$.
\begin{DLtt}{12345678}
\item[{\rm\bf Input parameters:}]
\item[PARAM] Character variable selecting the parameters.
\begin{DLtt}{12345678}
\item['MT,ALS'] Select the parameters  $\tm$ and $\alphahat_s$.
\item['MT,MH'] Select the parameters  $\tm$ and $\hm$.
\item['ALS,MH'] Select the parameters  $\hm$ and $\alphahat_s$.
\end{DLtt}
\item[{\rm\bf Output parameters:}]
\item[PAR1] Value of the first parameter.
\item[PAR2] Value of the second parameter.
\item[EPAR1] Value of the error of the first parameter.
\item[EPAR2] Value of the error of the second parameter.
\item[R12]  Value of the correlation coefficient between two parameters.
\item[CHI2] Value of the $\chi^2$ of the fit.
\end{DLtt}
\Example
\begin{XMP}
   CALL LTFIT2('MT,ALS',AMT,ALS,EMT,EALS,R12,CHI2)
\end{XMP}
 After this call the variable AMT contains the fitted top quark mass $\tm$,
 the variable ALS contains the fitted strong coupling constant
 $\alphahat_s( \zm )$, the variable EMT contains the error
 of the top quark mass,
 the variable EALS contains the error of the strong coupling constant,
the variable R12 contains the correlation coefficient between the errors of
 two parameters and
CHI2 contains the $\chi^2$ of the fit. \\ \\

\section{Sample program}

%\chapter*{APPENDIX C}
%{
%\newpage

%\begin{center}
%{\bf APPENDIX C}\\
%\vspace{0.5cm}
%\end{center}

\begin{verbatim}
      program example
      call ltdemo
      end

      SUBROUTINE LTDEMO
*----- template program to demonstrate LEPTOP package                *

*----- set all LEPTOP print flags off
      CALL LTFLAG('PRNT',0)

*----- initialisation
      CALL LTINIT(0)

*----- get some constants
      CALL LTGET('MZ',AMZ)
      CALL LTGET('GFERMI',GFERMI)
      CALL LTGET('ALBAR',ALBAR)

*----- get more constants
      CALL LTGET('MELE',amele)
      CALL LTGET('MMUO',ammuo)
      CALL LTGET('MTAU',amtau)
      CALL LTGET('MS',amstr)
      CALL LTGET('MC',amchrm)
      CALL LTGET('MB',ambot)

*----- modify some constants

      CALL LTPUT('MT',175.)
      CALL LTPUT('MH',300.)
      CALL LTPUT('ALSHAT',0.125)

*----- get some physical quantities
      CALL LTGET('GV',gv)
      CALL LTGET('GA',ga)
      CALL LTGET('MW',amw)
      CALL LTGET('GNU',gnu)
      CALL LTGET('GE',ge)
      CALL LTGET('GMUON',gmuon)
      CALL LTGET('GTAU',gtau)

*----- fit of mtop and alsbar
      CALL LTFIT2('MT,ALS',amt1,als1,eamt1,eals1,rho1,chi21)

*----- set data from Glasgow conference
      CALL LTFPUT('GLASGOW','ALL',0.)

*----- modify value of Z total width
      CALL LTFPUT('VALUE','GZ',2.4974)

*----- fit of mtop and mh
      CALL LTFIT2('MT,MH',amt2,amh1,eamt2,eamh1,rho2,chi22)

*----- fit  mt
      CALL LTFIT1('MT',amt3,eamt3,chi23)

*-----  fit  mh
      CALL LTFIT1('MH',amh4,eamh4,chi24)

*----- fit  als
      CALL LTFIT1('ALS',als,eals,chi2)

      END          ! LTDEMO
\end{verbatim}
%}

% Local Variables:
% mode: latex
% TeX-master: "hboomain"
% End:

\vspace{2cm}

\chapter*{Acknowledgments}

This description of LEPTOP was written as a backup document of
 our contribution to Electroweak
 Working Group Report \cite{202}.
 We are grateful to D.Yu.Bardin for many discussions and providing
us with useful information.

We are grateful to L.Avdeev,
K.G. Chetyrkin and A.L.Kataev for their comments on QCD corrections
; to V.P.Yurov and N.A.Nekrasov for their contributions to papers
\cite{3}, \cite{4}, \cite{8}; to V.L.Telegdi, without whose
insistence and help the flowchart would not appear; M.V. is grateful
to CPPM and CPT for kind hospitality at Marseille where this work was
finished; L.O., V.N. and M.V. are grateful to RFFR grant 93-02-14431
 and V.N.  and M.V.  are grateful to INTAS grant 93-3316 for partial
support.

\newpage
\chapter*{\bf APPENDIX A}

%\begin{center}
%{\bf APPENDIX A}\\
%\vspace{0.1cm}
%\end{center}
\begin{center}
{
\Huge
    FLOWCHART OF LEPTOP
}\\[3mm]
\setlength{\unitlength}{1mm}
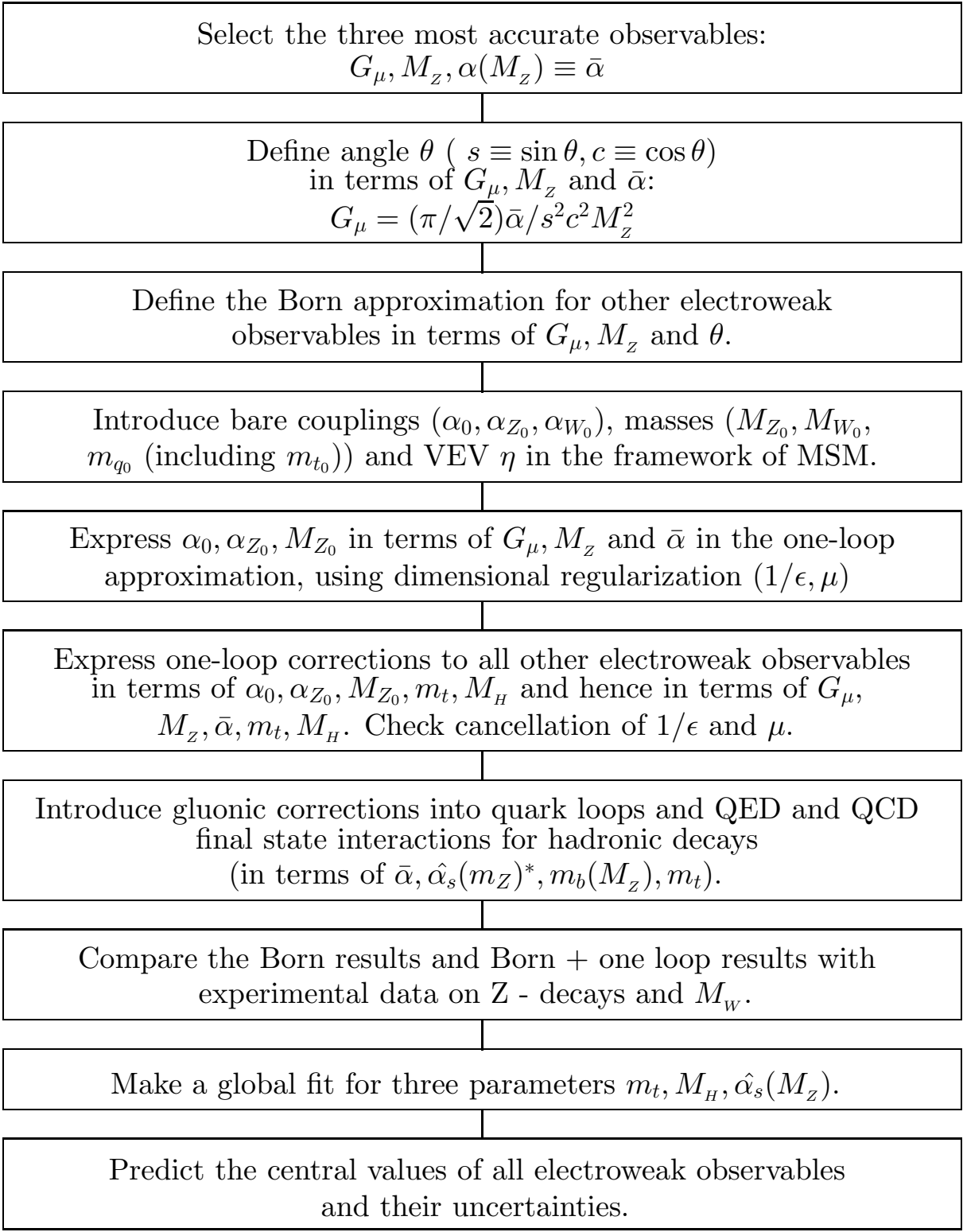
\begin{picture}(160,210)
\put(0,195){\framebox(160,15)}
\put(80,205){\makebox(0,0){\large
Select the three most accurate observables:}}
\put(80,199){\makebox(0,0){\large
$ \gf, \zm, \alpha(\zm) \equiv \bar{\alpha}$
}}
\put(80,190){\line(0,1){5}}

\put(0,170){\framebox(160,20)}
\put(80,185){\makebox(0,0){\large
Define angle $\theta$ ( $s \equiv \sin \theta, c \equiv \cos \theta$)}}
\put(80,180){\makebox(0,0){\large
in terms of $\gf, \zm$ and $\bar \alpha$:}}
\put(80,174){\makebox(0,0){\large
$\gf = (\pi / \sqrt{2}) \bar{\alpha} / s^2 c^2 \zm^2$}}
\put(80,165){\line(0,1){5}}

\put(0,150){\framebox(160,15)}
\put(80,160){\makebox(0,0){\large
Define the Born approximation for other electroweak}}
\put(80,154){\makebox(0,0){\large
observables in terms of $\gf, \zm$ and $\theta$.}}
\put(80,145){\line(0,1){5}}

\put(0,130){\framebox(160,15)}
\put(80,140){\makebox(0,0){\large
Introduce bare couplings ($\alpha_0, \alpha_{Z_0}, \alpha_{W_0}$),
masses ($M_{Z_0}, M_{W_0},$}}
\put(80,134){\makebox(0,0){\large
$ m_{q_0}$ (including $m_{t_0}$)) and VEV $\eta$
in the framework of MSM.}}
\put(80,125){\line(0,1){5}}

\put(0,110){\framebox(160,15)}
\put(80,120){\makebox(0,0){\large
Express $\alpha_0, \alpha_{Z_0},M_{Z_0}$
in terms of $\gf, \zm$ and $\bar \alpha$
in the one-loop}}
\put(80,114){\makebox(0,0){\large
approximation, using dimensional regularization ($1/ \epsilon, \mu$)
}}
\put(80,105){\line(0,1){5}}

\put(0,85){\framebox(160,20)}
\put(80,100){\makebox(0,0){\large
Express one-loop corrections to all other electroweak observables}}
\put(80,95){\makebox(0,0){\large
in terms of $\alpha_0, \alpha_{Z_0},M_{Z_0}, \tm , \hm$
and hence in terms of $\gf$,
}}
\put(80,89){\makebox(0,0){\large
$\zm,\bar \alpha, \tm, \hm$. Check cancellation of $1/ \epsilon$ and $\mu$.
}}
\put(80,80){\line(0,1){5}}

\put(0,60){\framebox(160,20)}
\put(80,75){\makebox(0,0){\large
Introduce gluonic corrections into quark loops and QED and QCD
}}
\put(80,70){\makebox(0,0){\large
                final state interactions for hadronic decays
}}
\put(80,64){\makebox(0,0){\large
(in terms of $\bar \alpha, \hat{\alpha_{s}}(m_Z)^*, m_b(\zm), \tm$).
}}
\put(80,55){\line(0,1){5}}

\put(0,40){\framebox(160,15)}
\put(80,50){\makebox(0,0){\large
Compare the Born results and Born + one loop results with
}}
\put(80,44){\makebox(0,0){\large
            experimental data on Z - decays and $\wm$.
}}
\put(80,35){\line(0,1){5}}

\put(0,25){\framebox(160,10)}
\put(80,29){\makebox(0,0){\large
Make a global fit for three parameters  $\tm, \hm, \hat{\alpha_s}(\zm)$.
}}
\put(80,20){\line(0,1){5}}

\put(0,5){\framebox(160,15)}
\put(80,15){\makebox(0,0){\large
Predict  the central values of all electroweak observables
}}
\put(80,9){\makebox(0,0){\large
                          and their uncertainties.
}}
\put(80,0){\makebox(0,0){\large
$* -  \hat{\alpha_s}(\zm)$ is the QCD coupling in $\overline{\mbox{\rm MS}}$
 - scheme.
}}
\end{picture}
\end{center}

\chapter*{APPENDIX B}
%\newpage

\begin{center}
%{\bf APPENDIX B}\\
%\vspace{0.5cm}
List of LEPTOP papers \footnote{V.Novikov, L.Okun, and M.Vysotsky
are coauthors of all fourteen papers} \end{center}

1. Electroweak radiative corrections and top quark mass, CERN-TH.6053/91,
March 1991, TPI-MINN-91/14-I, ITEP-15/91.

2. Parametrization of electroweak radiative corrections, ITEP-104/92,
November 1992; $ZhETF$ {\bf 103} (1993) 1489 (in Russian), May 1993, $JETP$
{\bf 76} (1993) 725.

3. On the electroweak one-loop corrections, CERN-TH.6538/92;  ITEP-67/92,
June 1992; Erratum August 1992; $Nucl.Phys.$ {\bf B397} (1993) 35.

4. On the interpretation of the CHARM II data, CERN-TH.6695/92, October
1992; $Phys.Lett.$ {\bf B298} (1993) 453.

5. Virtual gluons in the electroweak loops (with N.Nekrasov),
CERN-TH.6696/92, October 1992; $Yad.Fys.$ {\bf 57} (1994) 883.

6. Do-it-yourself analysis of precision electroweak data, CERN-TH.6715/92,
November 1992; Erratum March 1993; $Phys.Lett.$ {\bf B299} (1993) 329;
Erratum {\bf 304} (1993) 386.

7. The isolines of electroweak radiative corrections and the confidence
levels for the masses of the top and higgs (with V.Yurov), CERN-TH.6849/93,
March 1993; $Phys.Lett.$ {\bf B308} (1993) 123.

8. On the electroweak and gluonic corrections to the hadronic width of the
$Z$ boson, CERN-TH.6855/93, April 1993; $Phys.Lett.$ {\bf B320} (1994) 388.

9. Do present data provide evidence for electroweak corrections?
CERN-TH.6943/93, July 1993; $Modern Phys.Lett.$ {\bf A8} (1993) 2529;
Erratum {\bf 8} (1993) 3301.

10. On the effective electric charge in the electroweak theory,
CERN-TH.7071/93, November 1993; $Phys.Lett.$ {\bf B324} (1994) 89.

11. The values of $m_t$ and $\bar{\alpha}_s$ derived from non-observation of
electroweak radiative corrections at LEP: global fit (with A.Rozanov and
V.Yurov), CERN-TH.7137/94, January 1994; $Phys.Lett.$ {\bf B331} (1994) 433.

12. The $Q^2$ dependence of $W$ and $Z$ coupling constants in the interval
$0 \leq |q^2| \leq m^2_Z$, CERN-TH.7153/94, January 1994; $Modern
Phys.Letters$ {\bf A9} (1994) 1489.

13. First evidence for electroweak radiative corrections from the new
precision data (with A.Rozanov), CERN-TH.7217/94, April 1994; $Modern
Phys.Letters$ {\bf A9} (1994) 2641.

14. Do the present electroweak precision measurements leave room for extra
generations? (with A.Rozanov and V.Yurov), CERN-TH.7252/94, May 1994.

%\newpage

\end{document}